\documentclass[referee,a4paper,12pt,traditabstract]{swsc} 

\usepackage{graphicx}
\usepackage{txfonts}
\usepackage{subfigure}
\usepackage{epstopdf}
\usepackage[displaymath,mathlines]{lineno}
\usepackage[authoryear,round]{natbib}
\usepackage[backref]{hyperref}
\usepackage{url}
\usepackage{gensymb}

\hypersetup{colorlinks=true,citecolor=cyan,urlcolor=cyan,linkcolor=blue}

\begin{document}


   \title{Impact of solar magnetic field amplitude and geometry on cosmic rays diffusion coefficients in the inner heliosphere}

   \subtitle{}
   
   \titlerunning{Impact of magnetic field on cosmic rays}

   \authorrunning{Perri et al.}

   \author{B. Perri
          \inst{1}\fnmsep\inst{2},
          A. S. Brun
          \inst{2},
          A. Strugarek
          \inst{2},
          \and
          V. Réville\inst{3}
          }

   \institute{Université Paris-Saclay, CNRS,  Institut
              d'astrophysique spatiale, 91405, Orsay, France
         \and
             AIM, CEA, CNRS,
             Université Paris-Saclay, Université Paris-Diderot, Sorbonne Paris Cité,
             F-91191 Gif-sur-Yvette, France
        \and
             IRAP, Université de Toulouse, CNRS, UPS, CNES, Toulouse, France, 14 Avenue Edouard Belin, F-31400 Toulouse, France
             }
 
  \abstract
   {Cosmic rays are remarkable tracers of solar events when they are associated with solar flares, but also galactic events such as supernova remnants when they come from outside our solar system. Solar Energetic Particles (SEPs) are correlated with the 11-year solar cycle while Galactic Cosmic Rays (GCRs) are anti-correlated due to their interaction with the heliospheric magnetic field and the solar wind. 
   Our aim is to quantify separately the impact of the amplitude and the geometry of the magnetic field, both evolving during the solar cycle, on the propagation of cosmic rays of various energies in the inner heliosphere (within Earth orbit). We focus especially on the diffusion caused by the magnetic field along and across the field lines.
   To do so, we use the results of 3D magnetohydrodynamics (MHD) wind simulations running from the lower corona up to 1 AU. This gives us the structure of the wind and the corresponding magnetic field. The wind is modeled using a polytropic approximation, and fits and power laws are used to account for the turbulence.
   Using these results, we compute the parallel and perpendicular diffusion coefficients of the Parker cosmic ray transport equation, yielding 3D maps of the diffusion of cosmic rays in the inner heliosphere.
   By varying the amplitude of the magnetic field, we change the amplitude of the diffusion by the same factor, and the radial gradients by changing the spread of the current sheet. By varying the geometry of the magnetic field, we change the latitudinal gradients of diffusion by changing the position of the current sheets. By varying the energy, we show that the distribution of values for SEPs is more peaked than GCRs. For realistic solar configurations, we show that diffusion is highly non-axisymmetric due to the configuration of the current sheets, and that the distribution varies a lot with the distance to the Sun with a drift of the peak value. 
   This study shows that numerical simulations, combined with theory, can help quantify better the influence of the various magnetic field parameters on the propagation of cosmic rays. This study is a first step towards the resolution of the complete Parker transport equation to generate synthetic cosmic rays rates from numerical simulations.}       

   \keywords{MHD --
                solar wind --
                cosmic rays
               }

   \maketitle

\section{Introduction}

The Sun possesses a magnetic field that shows a cyclic evolution in time : it has a cycle of 11 years in amplitude and 22 years in polarity on average, the shortest cycle observed being of 9 years and the longest one of 14 years \citep{hathaway_solar_2015, brun_magnetism_2017}. During a minimum of activity, the solar magnetic field has the lowest amplitude of the cycle and its geometry is mostly dipolar ; during a maximum of activity, its amplitude is at its peak and its geometry is mostly quadrupolar \citep{derosa_solar_2012}. This magnetic field is generated inside the star via a dynamo loop \citep{moffatt_magnetic_1978, parker_solar_1993} and fills the whole heliosphere, including the Earth spatial environment \citep{owens_heliospheric_2013}.

The heliosphere is filled with the solar wind, a continuous flow of charged particles ejected from the Sun \citep{neugebauer_solar_1962}. It has a slow and a fast component at respectively 400 and 800 km/s at Earth orbit, which corresponds to 1 AU, and hence is transsonic and transalfvénic at this distance from the Sun \citep{mccomas_three-dimensional_2003}. It was first described using fluid dynamics \citep{parker_dynamics_1958}, then magnetism was taken into account \citep{weber_angular_1967, sakurai_magnetic_1985}. Multiple models of the solar wind have been designed, from empirical models \citep{wang_magnetic_1990, arge_improvement_2000} to MHD numerical simulations in 1D \citep{lionello_including_2001, suzuki_saturation_2013, pinto_multiple_2017}, 2D \citep{keppens_numerical_1999, matt_accretion-powered_2008, reville_effect_2015} or 3D \citep{toth_adaptive_2012, riley_inferring_2015, reville_role_2020}. The heating of the corona is modeled through a polytropic approximation \citep{reville_global_2017} or via Alfvén waves perturbations \citep{usmanov_three-fluid_2014}. The complete list of phenomena leading to this heating still eludes our understanding and is a huge current numerical challenge \citep{reville_role_2020}.

Cosmic rays (CRs) are highly energetic extra-terrestrial particles with energies between $10^2$ MeV to $10^{11}$ GeV ; they follow a power-law distribution, except for the low-energy part of the distribution \citep{reames_particle_1999, heber_cosmic_2006}. They can be emitted by the Sun during sudden events such as solar flares or coronal mass ejections ; in that case they are called Solar Energetic Particles (SEPs) and correspond to the low-energy part of the distribution (up to 1 GeV). They can also be emitted by sudden events out of our solar system such as gamma-ray bursts or supernova remnants ; in that case, they are called Galactic Cosmic Rays (GCRs) and correspond to the high-energy part of the distribution (from 1 GeV) \citep{shalchi_nonlinear_2009}. As they progress through the heliosphere, CRs are subject to an adiabatic cooling while interacting with both the heliospheric magnetic field and the solar wind, which change significantly their trajectory due to their turbulent fluctuations \citep{parker_scattering_1964, jokipii_cosmic-ray_1966}. CRs rate is thus influenced by the cyclic activity of the Sun : SEPs are correlated with solar activity because sudden solar events are more frequent at maximum of activity ; on the contrary, GCRs are anti-correlated with solar activity because 
the magnetic field at maximum of activity makes it harder for GCRs to penetrate the heliosphere \citep{snyder_solar_1963, heber_cosmic_2006}. There are also a certain number of disparities in the CR distribution. The Voyager missions have suggested the presence of a negative latitudinal gradient in the count rate of $>70$ MeV protons \citep{cummings_latitudinal_1987}. Ulysses has shown that SEPs have a North-South asymmetry linked to the one of the magnetic field and the wind \citep{mckibben_three-dimensional_1998, perri_simulations_2018}; it has also shown that GCRs have larger gradients in their spatial distribution at minimum than at maximum \citep{belov_latitudinal_2003}. The observed electron to proton ratios, also linked to the radial and latitudinal gradients, indicate that large particle drifts are occurring during solar minimum, but diminish significantly towards solar maximum \citep{heber_cosmic_2006}.

To describe the propagation of CRs, the most common approach is the statistical one using the Parker cosmic ray transport equation from \cite{parker_passage_1965}. One of the biggest challenges in this equation is to determine the diffusion tensor, especially its dependency in space, time and energy. For the diffusion parallel to the magnetic field lines, the quasi-linear theory (QLT) yields good results, especially when extended to take into account time-dependent and non-linear corrections \citep{jokipii_cosmic-ray_1966, goldstein_nonlinear_1976, bieber_proton_1994, droge_solar_2003}. For the diffusion perpendicular to the magnetic field lines however, QLT provides only an upper limit using a field line random walk (FLRW) description \citep{jokipii_cosmic-ray_1966, forman_cosmic-ray_1974, giacalone_transport_1999}. Various alternate approaches were tested, including the Taylor-Green-Kubo (TGK) \citep{taylor_motion_1922, green_brownian_1951, kubo_statistical-mechanical_1957, forman_velocity_1977} or the Bieber and Matthaeus (BAM) \citep{bieber_perpendicular_1997} formulations, but these methods systematically underestimates the perpendicular diffusion \citep{bieber_nonlinear_2004}. Finally, the best method to this day is the non-linear guiding center (NLGC) theory \citep{matthaeus_nonlinear_2003, bieber_nonlinear_2004, shalchi_nonlinear_2009} which provides the best agreement with both observations and simulations. This is due to the assumption that there is a decorrelation between the diffusive spread of the particle gyrocenters following field lines and the diffusive spread of those field lines, due to the transverse complexity of the magnetic field \citep{matthaeus_nonlinear_2003}. There have been some recent reformulations of this theory such as the Extended NLGC (ENLGC) theory by \cite{shalchi_extended_2006} to improve the slab contribution, or with the random ballistic decorrelation (RBD) interpretation \citep{ruffolo_random_2012} which helped further improve the theory, by matching simulations over a wider range of fluctuation amplitudes ; recently there was also the application of the Reduced MHD (RMHD) for astrophysics derived by \cite{oughton_anisotropy_2015, oughton_reduced_2017} and the UNLT (Unified Non-Linear Transport) by \cite{ shalchi_time-dependent_2017, shalchi_perpendicular_2020} as a unified theory between all the previous approaches.

In most theoretical studies about CR diffusion, prescriptions are used for the magnetic field and the solar wind ; this usually limits the applications to certain range of energy or spatial locations. However, thanks to MHD numerical simulations, it is possible to have global descriptions of these quantities even in complex configurations. This approach has already been used in the studies of \cite{luo_galactic_2013}, \cite{guo_galactic_2014}, \cite{wiengarten_generalized_2016} and \cite{kim_predicting_2020} to predict the variations of CRs in the complete heliosphere, coupled with semi-empirical CRs prescriptions. However, we will focus here only on the very inner heliosphere within Earth orbit, which means that we do not include the whole dynamics of GCRs coming from outside the solar system. Similar approach has been used by \cite{chhiber_cosmic-ray_2017} to study the diffusion coefficient in the case of a tilted dipole. However in this study, only the inclination of the dipole was changed, and only meridional cuts were shown. In this study, we are interested in the correlation with cyclic activity of CRs, and thus want to study separately the two variations of the magnetic field over a cycle : variation in amplitude and in geometry. Thus we will characterize the difference between a reference case being a dipole of weak amplitude called D1, and a dipole of strong amplitude called D10 and a quadrupole of weak amplitude called Q1. Finally we will use the 3D aspect of our simulations by using the same method with realistic configurations of minimum and maximum of activity using synoptic maps. The minimum of activity corresponds to October 1995 and the maximum to August 1999 (Carrington rotations of CR 1902 and CR 1954 respectively). Both maps come from Wilcox Observatory. All results of the simulations and corresponding post-processing are available at the MEDOC online facility.\footnote{\url{https://idoc.ias.u-psud.fr/wind\_predict\_cr/}}

The article is organized as follows. Section \ref{sec:model} presents the wind model used for the simulations and the post-processing statistical computation of the cosmic rays diffusion coefficient. Section \ref{sec:study} details the various parametric studies we performed on the impact of the magnetic field amplitude and geometry, and the cosmic rays energy. Section \ref{sec:config} introduces our results for realistic configurations corresponding to a minimum and maximum of activity. Finally section \ref{sec:concl} sums our conclusions on this study and perspectives for future work.

\section{Model and equations}
\label{sec:model}

In this section we present first the 3D MHD wind model we used to derive the magnetic field structure and intensity and the solar wind speed in the inner heliosphere. Then we present the model used in post-processing to compute the diffusion of cosmic rays along and across magnetic field lines.

\subsection{Wind model}
\label{subsec:model_wind}

Our wind model is adapted from \cite{reville_effect_2015, reville_global_2017} and \cite{perri_simulations_2018} using the multi-physics compressible PLUTO code \citep{mignone_pluto:_2007}. In these articles the code has shown good agreement with other wind codes such as the model from \cite{matt_accretion-powered_2008}, or the code DIP \citep{grappin_search_2010}; we have also checked for the conservation of MHD invariants in \cite{strugarek_numerical_2015} and \cite{reville_effect_2015}. We solve the set of the conservative ideal MHD equations composed of the continuity equation for the density $\rho$, the momentum equation for the velocity field $\mathbf{v}$ with its momentum written $\mathbf{m}=\rho\mathbf{v}$, the energy equation which is noted $E$ and the induction equation for the magnetic field $\mathbf{B}$:
\begin{equation}
\frac{\partial}{\partial t}\rho+\nabla\cdot\left(\rho\mathbf{v}\right)=0,
\end{equation} 
\begin{equation}
\frac{\partial}{\partial t}\mathbf{m}+\nabla\cdot(\mathbf{mv}-\mathbf{BB}+\mathbf{I}p) = \rho\mathbf{a},
\end{equation}
\begin{equation}
\frac{\partial}{\partial t}E + \nabla\cdot((E+p)\mathbf{v}-\mathbf{B}(\mathbf{v}\cdot\mathbf{B})) = \mathbf{m}\cdot\mathbf{a},
\end{equation}
\begin{equation}
\frac{\partial}{\partial t}\mathbf{B}+\nabla\cdot(\mathbf{vB}-\mathbf{Bv})=0,
\label{eq:induction_pluto_wind}
\end{equation}
where $p$ is the total pressure (thermal and magnetic), $\mathbf{I}$ is the identity matrix and $\mathbf{a}$ is a source term (gravitational acceleration in our case). We use a polytropic assumption, which yields the following ideal equation of state: $\rho\varepsilon = p_{th}/(\gamma -1)$,
where $p_{th}$ is the thermal pressure, $\varepsilon$ is the internal energy per mass and $\gamma$ is the adiabatic exponent. This gives for the energy : $E = \rho\varepsilon+\mathbf{m}^2/(2\rho)+\mathbf{B}^2/2$.

PLUTO solves normalized equations, using three variables to set all the others: length, density and speed. If we note with $*$ the parameters related to the star and with $0$ the parameters related to the normalization, we have $R_*/R_0=1$, $\rho_*/\rho_0=1$ and $v_{kep}/V_0=\sqrt{GM_*/R_*}/V_0=1$, where $v_{kep}$ is the Keplerian speed at the stellar surface and $G$ the gravitational constant. By choosing the physical values of $R_0$, $\rho_0$ and $V_0$, one can deduce all of the other values given by the code in physical units. In our set-up, we choose $R_0=R_\odot=6.96 \ 10^{10}$ cm, $\rho_0=\rho_\odot=1.67 \ 10^{-16} \ \mathrm{g/cm}^3$ and $V_0=v_{kep,\odot}=4.37 \ 10^2$ km/s. We can define the escape velocity from the Keplerian speed as $v_{esc} = \sqrt{2}v_{kep} = \sqrt{2GM_\odot/R_\odot}$.

\begin{table}[!t]
    \centering
    \begin{tabular}{c|cc}
        \hline
        Control parameter & PLUTO control parameters & Coronal parameters \\ \hline
        Density & $\rho_0=1.67 \times 10^{-16} \ \mathrm{g/cm}^3$ & $n=1.0\times10^8 \ \rm{cm}^{-3}$ \\
        Rotation rate & $v_{rot}/v_{esc} = 2.93 \times 10^{-3}$ & $\Omega_0=2.6\times10^{-6} \ \rm{s}^{-1}$ \\ 
        Temperature & $c_s/v_{esc}=0.243$ \& $\gamma=1.05$ & $1.6 \ 10^6 \ \rm{K}$ \\
        Magnetic amplitude & $v_A/v_{esc}=0.176$ & $0.5 \ \rm{G}$ \\ \hline
    \end{tabular}
    \caption{Table of the control parameters of the 3D MHD wind simulation for the reference case D1. The four control parameters are the density, rotation rate, temperature and magnetic amplitude. They are expressed in the PLUTO code normalization system and with the correspondence in physical values.}
    \label{tab:ref_case}
\end{table}

Our wind simulations are controlled by four parameters : the adiabatic exponent $\gamma$ for the polytropic wind, the normalized rotation of the star $v_{rot}/v_{esc}$, the normalized speed of sound $c_s/v_{esc}$ and the normalized Alfvén speed at the equator $v_A/v_{esc}$. For the rotation speed, we take the solar value, which gives $v_{rot}/v_{esc} = 2.93 \ 10^{-3}$. We define a reference case for our simulations named D1. For this case, we choose to fix $c_s/v_{esc}=0.243$, which corresponds to a $1.6 \ 10^6$ K hot corona for solar parameters and $\gamma=1.05$. This choice of $\gamma$ is dictated by the need to maintain an almost constant temperature as the wind expands, which is what is observed in the solar wind. Hence, choosing $\gamma \neq 5/3$ is a simplified way of taking into account heating in the low corona, which is not modeled here. However a new model based on heating by Alfvén waves has been developed and has shown good agreement with the Parker Solar Probe data, see \cite{reville_role_2020}. The amplitude of the magnetic field is set by $v_A/v_{esc}=0.176$, which corresponds to an amplitude of 0.5 G at the equator for a dipole. All these parameters are summed up in table \ref{tab:ref_case}. Some parameters will vary depending on the model discussed, the differences between the various cases are shown in table \ref{tab:study_cases}.

We use the spherical coordinates $(r,\theta,\phi)$. We choose a finite-volume method using an approximate Riemann Solver (here the HLL solver, cf. \cite{einfeldt_godunov-type_1988}). PLUTO uses a reconstruct-solve-average approach using a set of primitive variables $(\rho,\mathbf{v},p,\mathbf{B})$ to solve the Riemann problem corresponding to the previous set of equations. The time evolution is then implemented via a second order Runge-Kutta method. To enforce the divergence-free property of the field, we use a hyperbolic divergence cleaning, which means that the induction equation is coupled to a generalized Lagrange multiplier in order to compensate the deviations from a divergence-free field \citep{dedner_hyperbolic_2002}. We use a splitting between the curl-free background field and the deviation field.

\begin{figure}[!t]
    \centering
    \subfigure{\includegraphics[width=0.6\columnwidth]{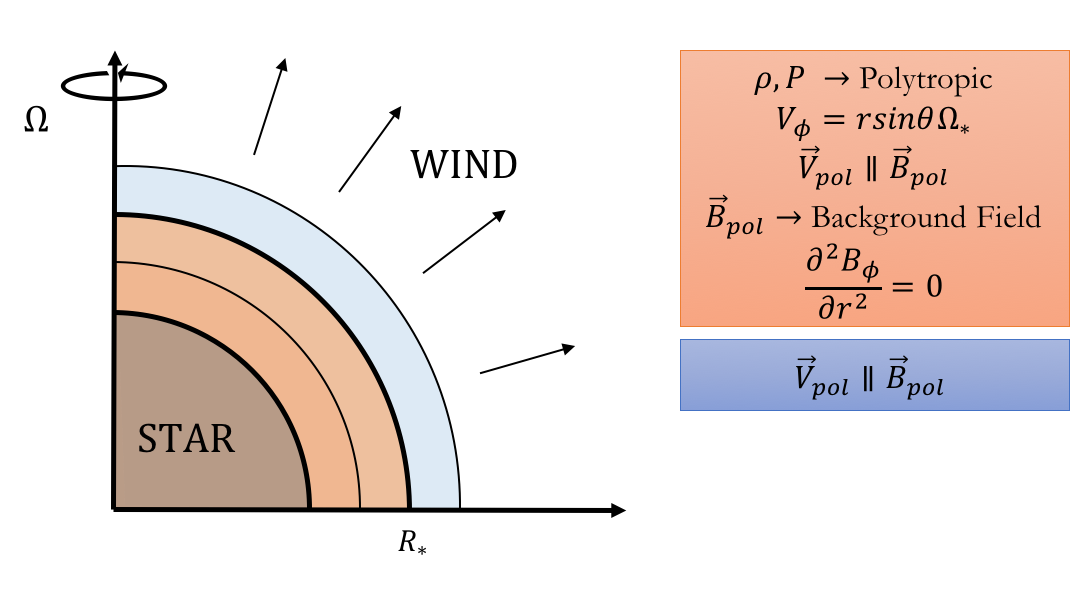}}
    \vspace{1cm}
    \subfigure{\includegraphics[width=0.3\columnwidth]{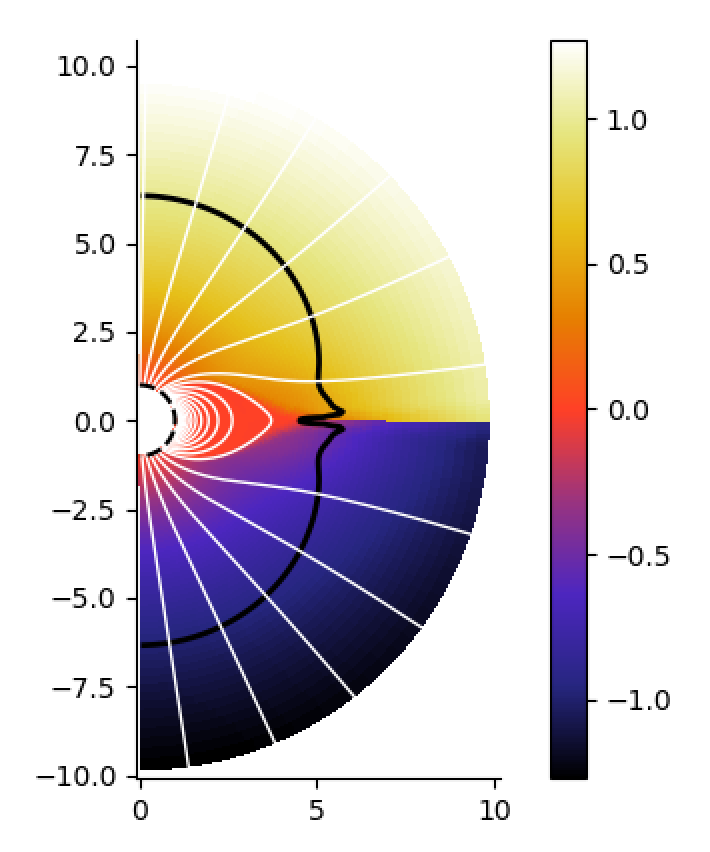}}
    \caption{Boundary conditions (on the left) and example of wind simulation (on the right). For the right panel, we show the relaxed state corresponding to the reference case D1. The color scale represents the following quantity : $\mathbf{v}\cdot\mathbf{B}/(c_s||\mathbf{B}||)$, which is the solar wind velocity projected on the magnetic field in units of Mach number. The black line corresponds to the Alfvén surface where the wind speed equals the Alfvén speed. White lines correspond to the poloidal magnetic field lines of positive polarity in solid and negative polarity in dashed lines. We represent only the 10 first solar radii.}
    \label{fig:wind_model}
\end{figure}

The numerical domain dedicated to the wind computation is a 3D sphere with the radius $r \in [1.001,220]R_\odot$, the co-latitude $\theta \in [0,\pi]$ and the longitude $\phi \in [0,2\pi]$. We use an uniform grid in latitude and longitude with respectively 256 and 512 points, and a stretched grid in radius with 256 points; the grid spacing is geometrically increasing from $\Delta r/R_\odot=0.001$ at the surface of the star to $\Delta r/R_\odot=0.02$ at the outer boundary. At the latitudinal boundaries ($\theta=0$ and $\theta=\pi$), we set axisymmetric boundary conditions. At the longitudinal boundary conditions ($\phi=0$ and $\phi=2\pi$), we set periodic boundary conditions. At the top radial boundary ($r=220 R_\odot$), we set an outflow boundary condition which corresponds to $\partial/\partial r=0$ for all variables, except for the radial magnetic field where we enforce $\partial(r^2B_r)/\partial r=0$. Because the wind has opened the field lines and under the assumption of axisymmetry, this ensures the divergence-free property of the field. The bottom boundary conditions are shown in the left panel of figure \ref{fig:wind_model}. In the ghost cells (in orange), the density $\rho$ and pressure $p$ are set to a polytropic profile, the rotation is uniform, the poloidal speed $V_{\rm{pol}}$ is aligned with the poloidal magnetic field $B_{\rm{pol}}$; the latter is imposed by a background dipolar field while the toroidal magnetic field $B_\phi$ is linear. In the first point of the computational domain (in blue), all physical quantities are free to evolve, except for the poloidal speed $V_{\rm{pol}}$ which is forced to be aligned with the poloidal magnetic field $B_{\rm{pol}}$ to minimize the generation of currents at the surface of the star and keep it as close as possible to a perfect conductor. We initialize the velocity field with a 1D Parker-like polytropic wind solution and the magnetic field with either a dipole, a quadrupole or a realistic magnetic field configuration from a synoptic map depending on the study case.

\subsection{Cosmic-ray post-processing}
\label{subsec:model_cr}

\subsubsection{Diffusion coefficients}
\label{subsubsec:diffusion_coef}

We use a statistical approach based of a Fokker-Planck equation, starting from the Parker cosmic ray transport equation presented in \cite{parker_passage_1965}:
\begin{equation}
\frac{\partial U}{\partial t} + \frac{\partial}{\partial x_i}\left(Uv_i\right) + \frac{\partial}{\partial T} \left(U\frac{\partial T}{\partial t}\right) - \frac{\partial}{\partial x_i}\left(\kappa_{ij}\frac{\partial U}{\partial x_{j}}\right) = 0,
\label{eq:cr_prop_parker}
\end{equation}
where  $U(x_i,T,t)$ is the distribution of cosmic rays depending on their spatial coordinates $x_i$, kinetic energy $T$ and time $t$, $\kappa_{ij}$ is the diffusion tensor and $v_i$ is the wind speed.
With the numerical simulations using the wind model described above, we have a prescription for the wind speed and the magnetic field structure and amplitude ; the only term left to compute is the diffusion tensor $\kappa$ which needs to be modeled. We will focus on this aspect in this article.

The diffusion tensor can be decomposed into three terms \citep{jokipii_convection_1970}:
\begin{equation}
\kappa_{ij} = \kappa_{\perp}\delta_{ij} + \frac{B_iB_j}{B^2}(\kappa_{||} - \kappa_{\perp}) + \epsilon_{ijk}\kappa_A\frac{B_k}{B},
\end{equation} 
with $\delta_{ij}$ being the Kronecker symbol and $\epsilon_{ijk}$ the Levi-Civita tensor. $\kappa_\parallel$ is the diffusion along the magnetic field lines and $\kappa_\perp$ is the diffusion across the magnetic field lines. The coefficient $\kappa_A$ is the drift coefficient and intervenes mostly for very energetic particles and strong gradients of the magnetic field \citep{jokipii_effects_1977}. It takes into account the influence of the current sheet \citep{jokipii_effects_1981}, the solar tilt angle \citep{lockwood_intensities_2005} and heliospheric magnetic field polarity \citep{webber_time_2005}. Drift effects also contribute to the 22-year cycle observed in the CR intensity and the CR latitudinal gradients \citep{heber_spatial_1996}. Recent studies have shown that the drift coefficient needs to be reduced to match spacecraft observations due to turbulence \citep{manuel_time-dependent_2011}, especially for SEPs in the inner heliosphere \citep{engelbrecht_comparison_2015, engelbrecht_toward_2017}. In this study, we will first focus on the diffusion coefficients which are better characterized and more prominent for SEPs in the inner heliosphere, and we will include drift effects in a later study. We can also introduce mean free path (mfp) $\lambda$, related to the diffusion tensor $\kappa$ by the following relationship:
\begin{equation}
\kappa_\parallel = \frac{1}{3}\lambda_\parallel \mathbf{v}_{CR}, \kappa_\perp = \frac{1}{3}\lambda_\perp \mathbf{v}_{CR},
\end{equation}
where $v_{CR}$ is the particle speed.

In this study, we will thus focus on describing the parallel and perpendicular mean free paths (mfps). To do so, we will not go into details into all the formulations proposed, but only used the most recent ones which have reached a general consensus, detailed hereafter; the reader can however find some very complete reviews in \cite{shalchi_nonlinear_2009} for parallel diffusion and \cite{shalchi_perpendicular_2020} for perpendicular diffusion. The most efficient geometry to describe such parameters is the composite geometry described in \cite{bieber_proton_1994} with 80\% of 2D geometry (both magnetic fluctuations and wave vectors are perpendicular to the magnetic field) and 20\% of slab geometry (magnetic fluctuations are perpendicular to the magnetic field, but wave vectors are parallel to it). It is also supported by wind observations that show a strong 2D component of the turbulence \citep{matthaeus_evidence_1990}. From now on, quantities related to the 2D geometry will be noted with an index $2$ while quantities related to the slab geometry will be noted with an index $s$.

A good approximation for parallel mfp is given by \cite{zank_radial_1998}:
\begin{equation}
\lambda_{\parallel} = 6.2742\frac{B^{5/3}}{\langle b_s^2\rangle}\left(\frac{P}{c}\right)^{1/3}\lambda_s^{2/3}\left[1+\frac{7A_\parallel/9}{(1/3+q)(q+7/3)}\right] \ \rm{km},
\label{eq:lambda_par}
\end{equation}
with $B$ the magnetic field norm, $\langle b_s^2\rangle$ the variance of the slab geometry fluctuation, $P = \tilde{p}c/Ze$ the particle rigidity ($\tilde{p}$ being the particle momentum and $Ze$ the particle charge), $c$ the speed of light, $\lambda_s$ the correlation length for the slab turbulence, and:
\begin{equation}
A_\parallel = (1+s^2)^{5/6}-1,
\end{equation}
\begin{equation}
q = \frac{5s^2/3}{1+s^2-(1+s^2)^{1/6}},
\end{equation}
and :
\begin{equation}
s = 0.746834\frac{R_L}{\lambda_s},
\end{equation}
with $R_L=P/Bc$ being the particle Larmor radius. The units are specific in this formula : as explained in \cite{bieber_diffusion_1995}, the magnetic field $B$ is in nT, the magnetic fluctuations $\langle b_s^2\rangle$ in $\rm{nT}^2$, the rigidity $P$ in V, the light speed $c$ in m/s, the correlation length $\lambda_s$ in m, and the final mfp in km. It is valid for rigidities ranging from 10 MV to 10 GV. We use the formula $E=\sqrt{(PZ/A)^2+E_0^2}-E_0$ to go from rigidity to energy, where $Z$ is the particle charge and $A$ is the mass number. In this study we will consider only protons with $Z=1$ and $A=1$. Thus the rigidity range between 10 MV and 10 GV is equivalent to energies ranging between 53 keV and 9 GeV.

For the perpendicular diffusion, the formulation that best fits both observations and data is the nonlinear guiding center (NLGC) theory described in \cite{bieber_nonlinear_2004}. In \cite{shalchi_analytic_2004}, analytical forms were derived from NLGC depending on the rigidity of the particle. This formulation also presents the advantage that the perpendicular diffusion only depends on the parallel diffusion and the magnetic field properties. Hence, for small rigidities ($P<10^2$ MV):
\begin{equation}
    \lambda_\perp \approx \frac{a^2}{2}\frac{\langle b_2^2\rangle}{B^2}10^{-3}\lambda_\parallel,
\end{equation}
with $a^2=1/3$ is a numerical factor determined by simulations \citep{matthaeus_nonlinear_2003}. 

For high rigidities ($P>10^2$ MV):
\begin{equation}
    \lambda_\perp \approx \left(\frac{2\nu-1}{4\nu}F_2(\nu)\lambda_sa^2\frac{\langle B'^2\rangle}{B^2}\frac{2\sqrt{3}}{25}\right)^{2/3}\lambda_\parallel^{1/3} \ \rm{m},
\end{equation}
with the spectral index $\nu=5/6$ and $F_2(\nu)=\sqrt{\pi}\frac{\Gamma(\nu+1)}{\Gamma(\nu+1/2)}$.

In these two formulations, $\langle b_2^2\rangle$ is in $\rm{nT}^2$, $B$ is in nT, $\lambda_\parallel$ is in km, $\lambda_s$ is in m and the resulting $\lambda_\perp$ is in m.

\subsubsection{Modeling the turbulence}
\label{subsubsec:turb}

As the wind model chosen here does not solve yet any equation linked to the turbulence (see \cite{reville_role_2020} for a first step in this direction), we need to model the quantities $\langle b_s^2\rangle$ and $\lambda_s$. We will use the model described in \cite{chhiber_cosmic-ray_2017}.

To estimate $\langle b_s^2\rangle$, we combine the two following formulae linked to the composite geometry used:
\begin{equation}
\frac{\langle b_s^2\rangle}{\langle b_2^2\rangle} = \frac{1}{4}, \langle b_2^2\rangle + \langle b_s^2\rangle = \langle B'^2\rangle \Rightarrow \langle b_s^2\rangle = \frac{1}{5}\langle B'^2\rangle, \langle b_2^2\rangle = \frac{4}{5}\langle B'^2\rangle,
\end{equation}
and use the expression of $\langle B'^2\rangle$:
\begin{equation}
\langle B'^2\rangle = \frac{Z'^2}{M_A'+1}4\pi\rho,
\end{equation}
with $Z'^2 = \langle v'^2+b'^2\rangle$ being the fluctuation energy and $M_A' = \langle v'^2\rangle/\langle b'^2\rangle$ the Alfvén ratio. From observations \citep{tu_magnetohydrodynamic_1995}, we can approximate $M_A'$ by 1 if $r<45R_\odot$ and 1/2 beyond $45 R_\odot$. To model $Z'^2$, we can use the Alfvén wave energy density $\epsilon$ whose expression is $Z'^2 = 2\epsilon/\rho$. We finally obtain:
\begin{equation}
    \langle b_s^2\rangle = \frac{8}{5}\frac{\epsilon\pi}{M_A'+1},
    \langle b_2^2\rangle = \frac{32}{5}\frac{\epsilon\pi}{M_A'+1}.
\end{equation}

For the Alfvén wave energy density $\epsilon$, we needed an expression that could adapt to any amplitude and geometry of the magnetic field, as we wanted to vary these parameters. Instead of solving WKB computations for each case (similar to \cite{usmanov_global_2000}), we chose to perform a fit, using an Alfvén-wave turbulence model \citep{reville_role_2020}. This model propagates parallel and anti-parallel Alfvén waves following the WKB theory inside a MHD wind model similar to what has been described in the previous section. We can then reconstruct $\epsilon$ using the Elsässer variables $z^+$ and $z^-$:
\begin{equation}
    \epsilon = \frac{1}{4}\left((z^+)^2 + (z^-)^2)\right)\rho.
\end{equation}
Using the output from this model, we performed a fit of $\epsilon$ depending on several other physical quantities of the model. According to our tests, the most relevant quantities are the amplitude of the total magnetic field $B$, the current $J$ and the poloidal speed $v_p$. We obtained the following formula as best fit:
\begin{equation}
    \epsilon \approx 0.0104 \ B^{1.25} \ J^{0.166} \ v_p^{-0.123},
    \label{eq:epsilon_fit}
\end{equation}
with $B$, $J$ and $v_p$ in PLUTO code units, so normalized by respectively $B_0=\sqrt{4\pi\rho_0}V_0$ and $V_0=v_{kep}$; the final result $\epsilon$ is also in code units, and then normalized to fit the amplitude described in \cite{usmanov_global_2000} in $\rm{erg}.\rm{cm}^{-3}$. The standard deviation error associated with each parameter is respectively $1.73\times10^{-5}$, $2.47\times10^{-4}$, $2.25\times10^{-4}$ and $4.01\times10^{-4}$, which is reasonable given the parameter values. This fit was obtained for solar parameters, which means it is valid for magnetic fields between 0.1 and 20 G and for wind speeds between 300 and 800 km/s. We have not tested this fit for other ranges of input parameters.

\begin{figure}
    \centering
    \subfigure{\includegraphics[width=0.30\columnwidth]{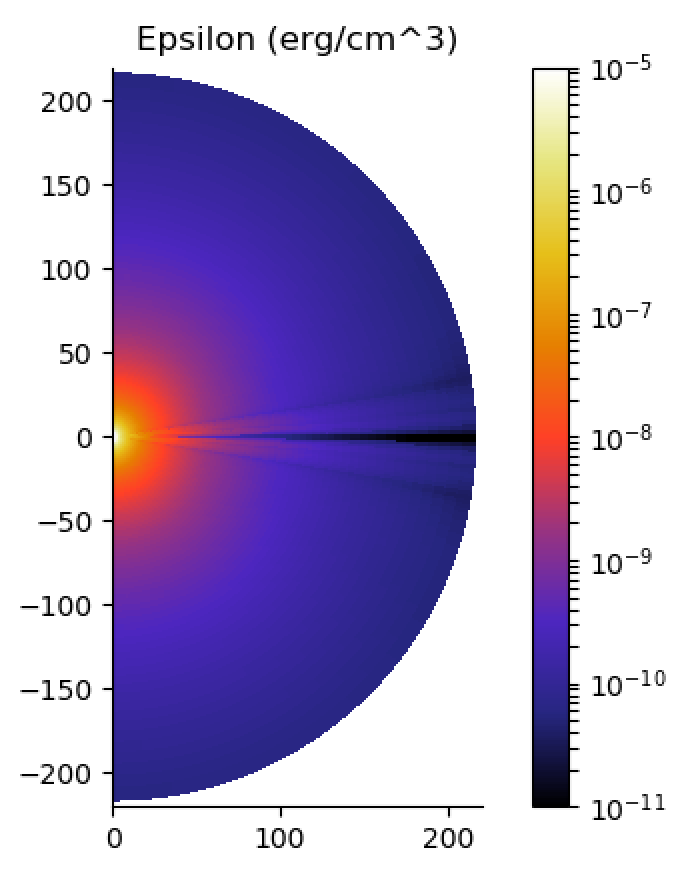}}
    \subfigure{\includegraphics[width=0.65\columnwidth]{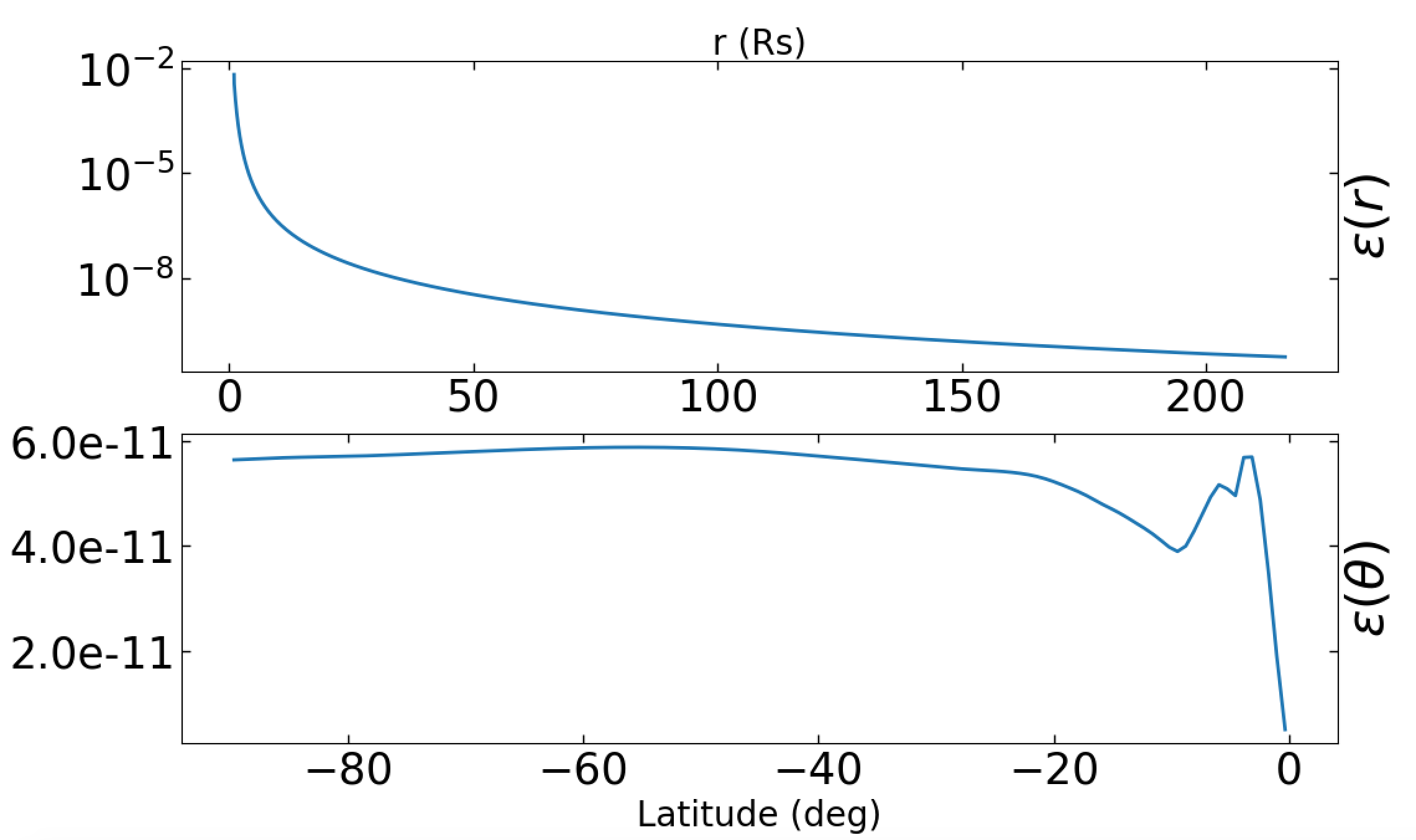}}
    \caption{2D (on the left) and 1D (on the right) representation of the Alfvén wave energy density $\epsilon$ obtained with our fit. For the left panel, we illustrate the profile of $\epsilon$ for the reference case D1 (see table \ref{tab:study_cases}). For the right panel, we show the radial profile taken at the north pole (top) and the latitudinal profile taken in the northern hemisphere at 1 AU. Values are in $\rm{erg}.\rm{cm}^{-3}$.}
    \label{fig:epsilon}
\end{figure}

Thus we can apply relation \ref{eq:epsilon_fit} to any case and have a turbulence which is self-consistent with the wind simulation. We show in figure \ref{fig:epsilon} an example of relation \ref{eq:epsilon_fit} applied to the reference case. The left panel shows the 2D meridional average of $\epsilon$ in $\rm{erg}.\rm{cm}^{-3}$ from the solar surface up to 1 AU. The right panel shows the radial profile taken at the north pole and the latitudinal profile in the northern hemisphere at 1 AU. Values are very similar to what was found in the case of \cite{usmanov_global_2000} for a dipole. The fact that we take into account the current $J$ however introduces a new effect at the borders of the current sheet, as seen in the latitudinal profile, with a drop in $\epsilon$ of about 30\% near the equator.

To estimate $\lambda_s$, we use the assumption from \cite{hollweg_transition_1986} which implies that the correlation length varies as the distance between the magnetic field lines, which in turn depends on the magnetic field amplitude \citep{spruit_magnetic_1981}, so that $\lambda \propto B^{-1/2}$. Then we use the relationship $\lambda_s = 2\lambda_2 = 2\lambda$, observed in the simulations of \cite{usmanov_three-fluid_2014}. Finally :
\begin{equation}
\lambda_s = 4\times10^7B^{-1/2} \ \rm{m}.
\end{equation}
The proportionality constant is set depending on the boundary conditions to reach a value of $2\times10^7$ m at the surface of the Sun. 

To validate the values provided by these formulae, we have both observational and computational points of comparison. The observations regarding mfps of cosmic rays are summarized under the Palmer consensus \citep{palmer_transport_1982} : the values of $\lambda_\parallel$ at 1 AU range between 0.08 and 0.3 AU for rigidities between $0.5$ MV and $5$ GV. For perpendicular diffusion, we can combine the observations from \cite{chenette_observations_1977} for Jovians electrons and \cite{burger_rigidity_2000} for Ulysses measurements of galactic protons : this yields values for $\lambda_\perp$ between 0.003 and 0.01 AU for rigidities between 1 MV and 10 GV. Numerical simulations have also been performed in \cite{bieber_nonlinear_2004} for the perpendicular diffusion. Finally another study described in \cite{chhiber_cosmic-ray_2017} has used similar methods to derive parallel and perpendicular mfps for a tilted dipole of amplitude 16 G at the poles. We discuss in the next section this comparison for various magnetic field configurations.

\section{Parametric study}
\label{sec:study}

Now that we have a model to compute the parallel and perpendicular mfps from simulations of a magnetic wind, we will apply this to a parametric study to understand how various magnetic configurations influence CR diffusion coefficient. To do so we will focus on 3 cases, described in table \ref{tab:study_cases}. Case D1 is the reference case, which parameters were shown in table \ref{tab:ref_case}, with a dipole of amplitude $B_*$ of 0.5 G (taken at the surface of the star at the equator). Case D10 is the same but with a dipole of amplitude $B_*$ of 5 G, so 10 times stronger than D1. Case Q1 is of the same equatorial amplitude as D1 but with a quadrupolar geometry instead of dipolar. This will allow us to study separately the effects of amplitude and geometry on the diffusion of CRs. Finally we will discuss the influence of energy to differentiate SEPs from GCRs. Because we study in this section only axisymmetric configurations, we will only focus on the meridional plane corresponding to an azimuthal average. 

\begin{table}[!h]
    \centering
    \begin{tabular}{c|c|c|c}
        \hline
        Magnetic parameters & D1 & D10 & Q1 \\ \hline
        Geometry & Dipole & Dipole & Quadrupole \\
        Amplitude (equator) & 0.5 G & 5 G & 0.5 G \\ \hline
    \end{tabular}
    \caption{Magnetic field parameters for the 3 cases D1, D10 and Q1 used in the parametric study. The case D1 corresponds to a dipole of amplitude 0.5 G, the case D10 to a dipole of amplitude 5 G and the case Q1 to a quadrupole of amplitude 0.5 G. The amplitude is specified at the surface of the star at the equator. For the other physical parameters see table \ref{tab:ref_case}.}
    \label{tab:study_cases}
\end{table}

\subsection{Influence of amplitude}
\label{subsec:study_ampli}

Over an activity cycle, the surface magnetic field amplitude varies between minima and maxima of activity, typically by a factor between 4 and 10 \citep{derosa_solar_2012}. In order to characterize the impact of a variation of amplitude, we study how the increase by a factor 10 of the whole amplitude impacts the diffusion of CRs. To do so, we will focus on cases D1 and D10 described in table \ref{tab:study_cases}. We will focus here on protons of rigidity 445 MV, which corresponds to 100 MeV energy, for comparison with previous studies \citep{pei_cosmic_2010, chhiber_cosmic-ray_2017}.  

\begin{figure}[!t]
    \centering
    \subfigure{\includegraphics[width=0.45\columnwidth]{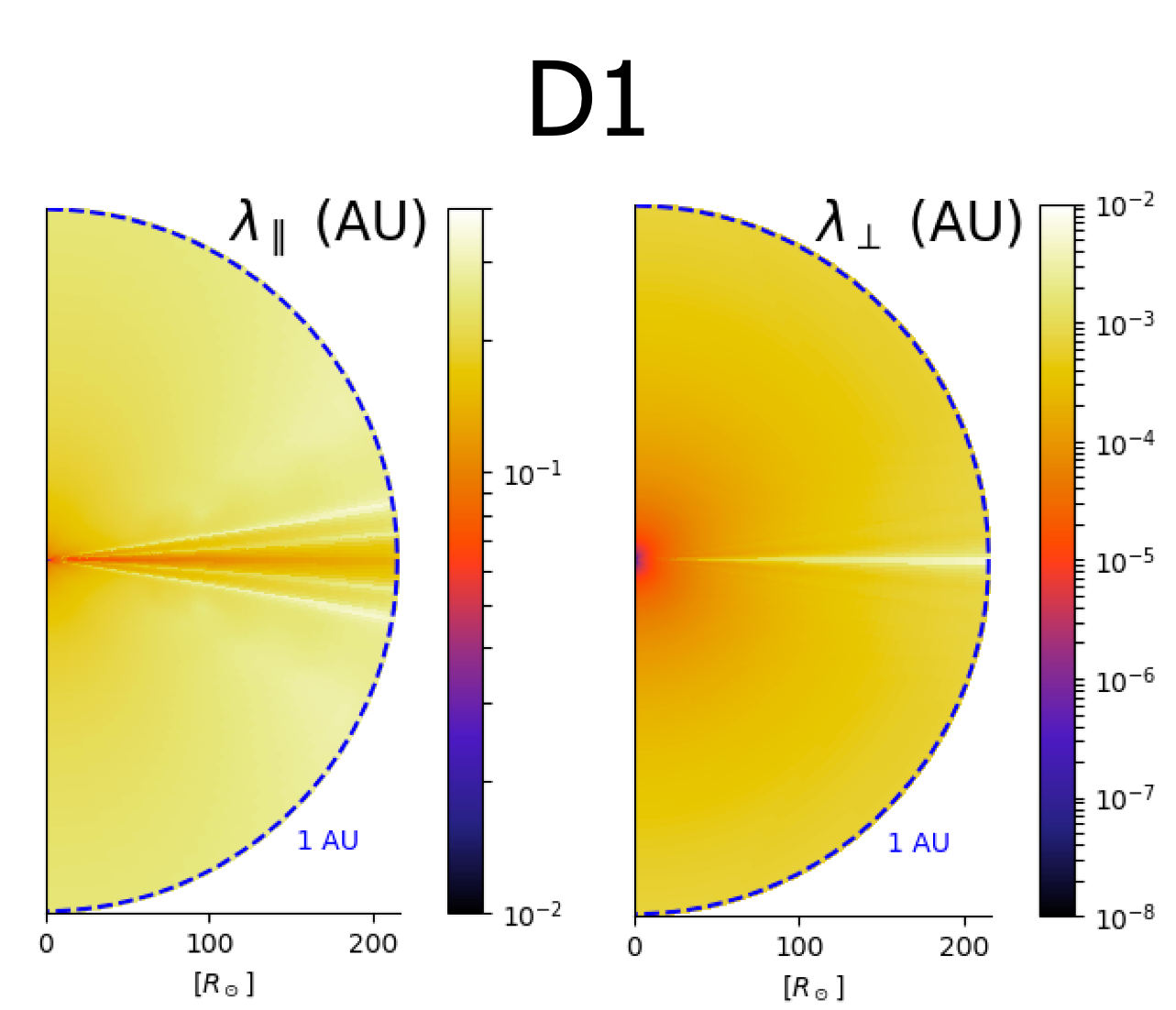}}
    \subfigure{\includegraphics[width=0.45\columnwidth]{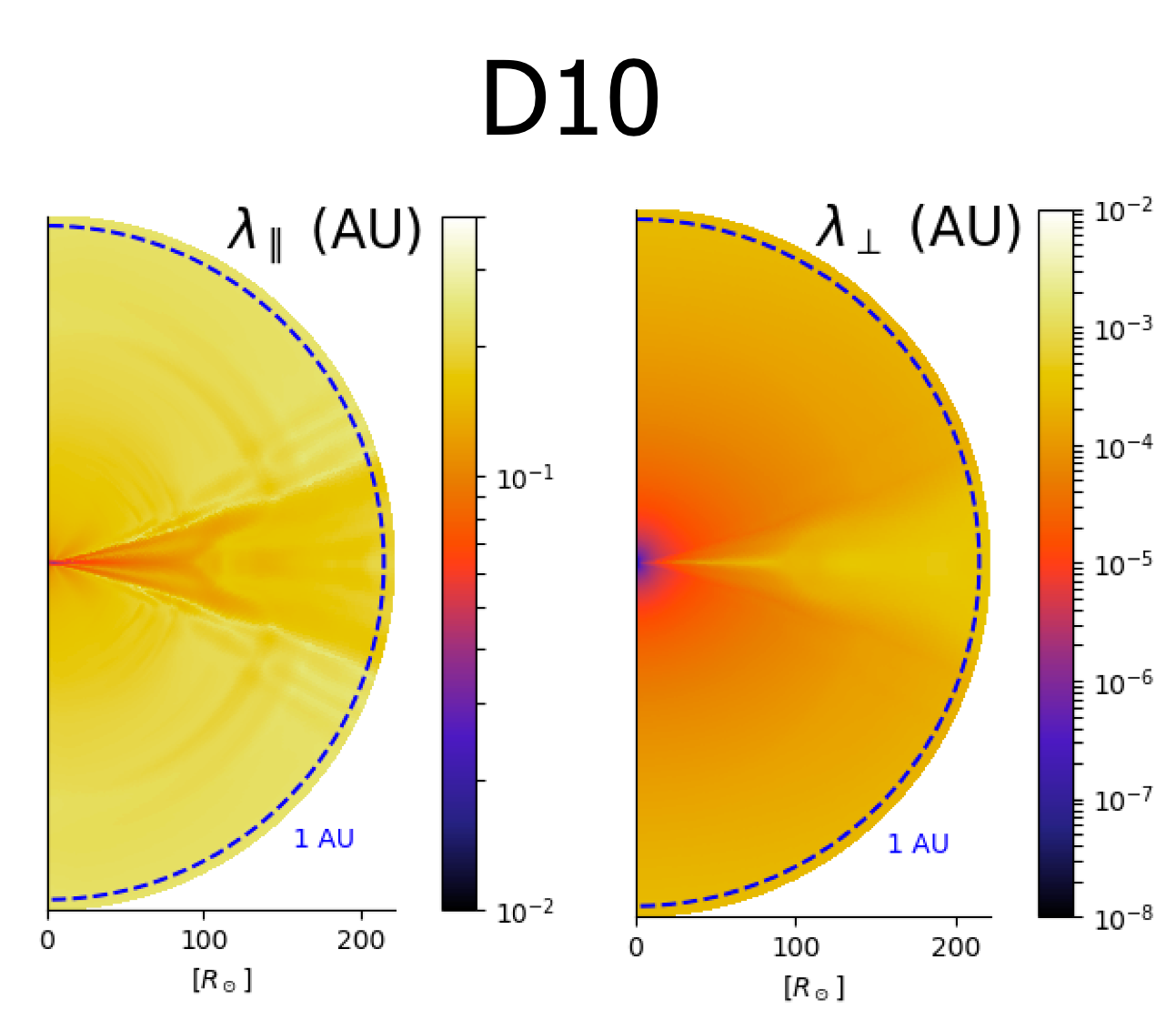}}
    \caption{Meridional cuts of the parallel mfp $\lambda_\parallel$ and the perpendicular mfp $\lambda_\perp$ in AU for a proton of rigidity 445 MV (which means 100 MeV energy). The two panels on the left correspond to the case D1, the two panels on the left correspond to the case D10. The same color scale is used for each mfp for both cases.}
    \label{fig:ampli_2d}
\end{figure}

Figure \ref{fig:ampli_2d} shows the 2D meridional cuts of the parallel and perpendicular mfps in units of AU ; we recall that 1 AU is equal to approximately $1.5\times 10^8$ km. On the left we have the case D1 and on the right the case D10. We used the same color scale for each mfp, independently of the case studied, to allow better comparison between the figures. 
We see first that the two mfps behave differently: the perpendicular mfp is nearly constant in the domain while the parallel mfp increases with distance. We find that for case D1, $\lambda_\parallel$ varies between 0.05 AU and 0.2 AU depending on the latitude (which corresponds to $4.5\times10^7$ km), while $\lambda_\perp$ varies between $10^{-8}$ AU close to the star in the low corona to $10^{-3}$ AU at Earth orbit (between 1.5 and $1.5\times10^5$ km).
They also have opposite behaviors at the heart of current sheets (see equatorial plane) : $\lambda_\parallel$ decreases while $\lambda_\perp$ increases \citep{bieber_nonlinear_2004, chhiber_cosmic-ray_2017}. Finally we can notice the same phenomenon on the edges of the current sheet : this is where $\lambda_\parallel$ actually reaches its highest value and where $\lambda_\perp$ loses one order of magnitude ; however, it seems that $\lambda_\parallel$ is more sensitive to the edges of the current sheet than $\lambda_\perp$, as we see more structures with better contrast. Since the NLGC formulation implies that $\lambda_\perp$ is proportional to $\lambda_\parallel$, it is expected to find similar structures for the two mfps.

From the 2D figure, we can see clearly that the amplitude of the stellar magnetic field has an effect near the equator on the spread of the current sheet. In case D1, the mfps are affected at the equator at 1 AU between $\theta=87\degree$ and $\theta=93\degree$ with $\lambda_\parallel$ being decreased by 75\% of its magnitude from 0.3 AU to 0.08 AU, and $\lambda_\perp$ being increased by 2 orders of magnitude from $10^{-4}$ AU to $10^{-2}$ AU. We recall that in ideal MHD such as for this model, the thickness of the current sheet is determined by the numerical resistivity, which naturally increases with distance in the case of a stretched grid like we have ; this implies that the expansion of the current sheet with the distance is most likely a numerical effect. In case D10, the zone of influence of the current sheet is now from $\theta=75\degree$ to $\theta=105\degree$ at 1AU, which is 5 times bigger. The rest of the numerical domain is isotropic.

\begin{figure}[!t]
    \centering
    \subfigure{\includegraphics[width=0.45\columnwidth]{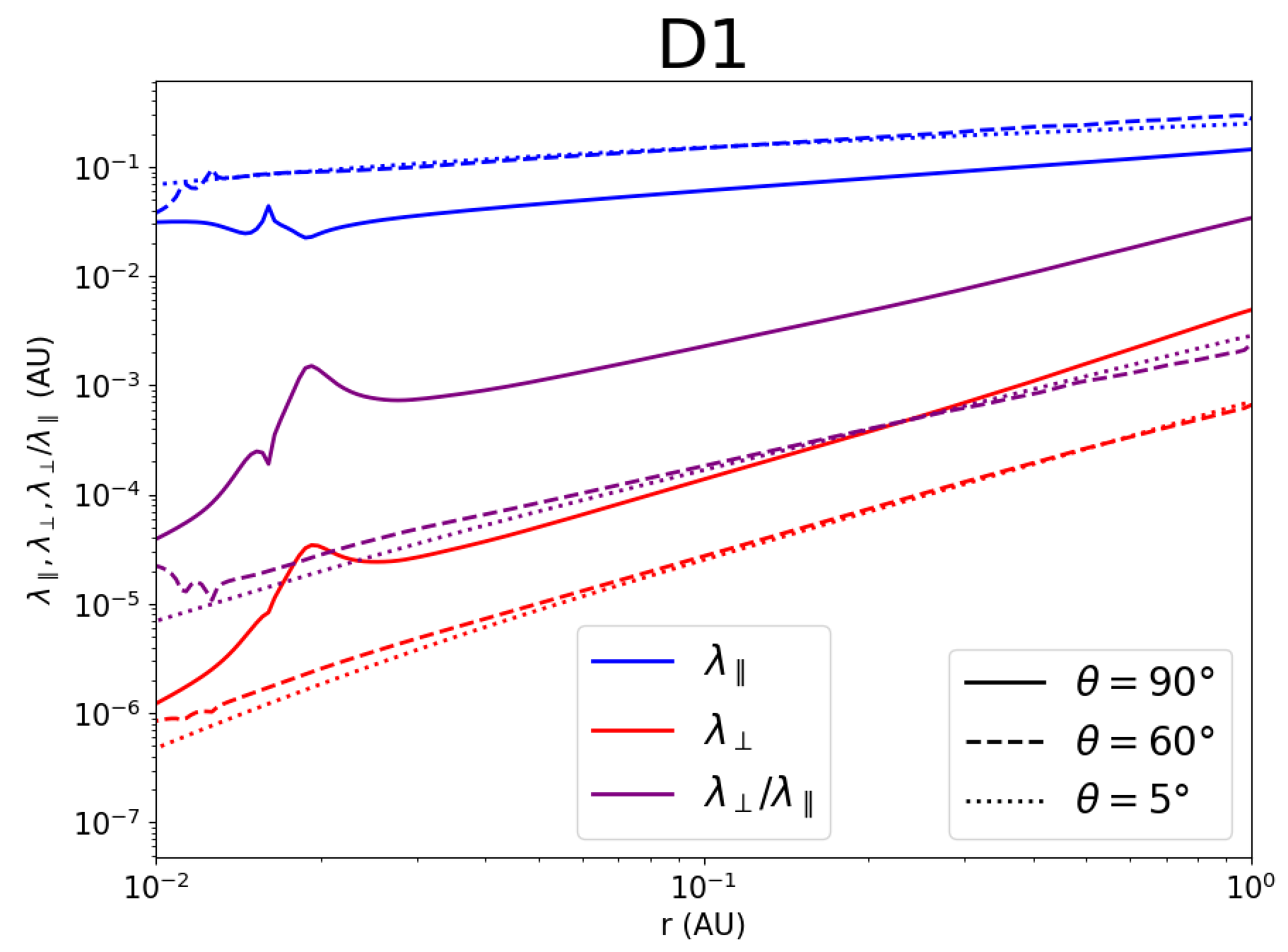}}
    \subfigure{\includegraphics[width=0.45\columnwidth]{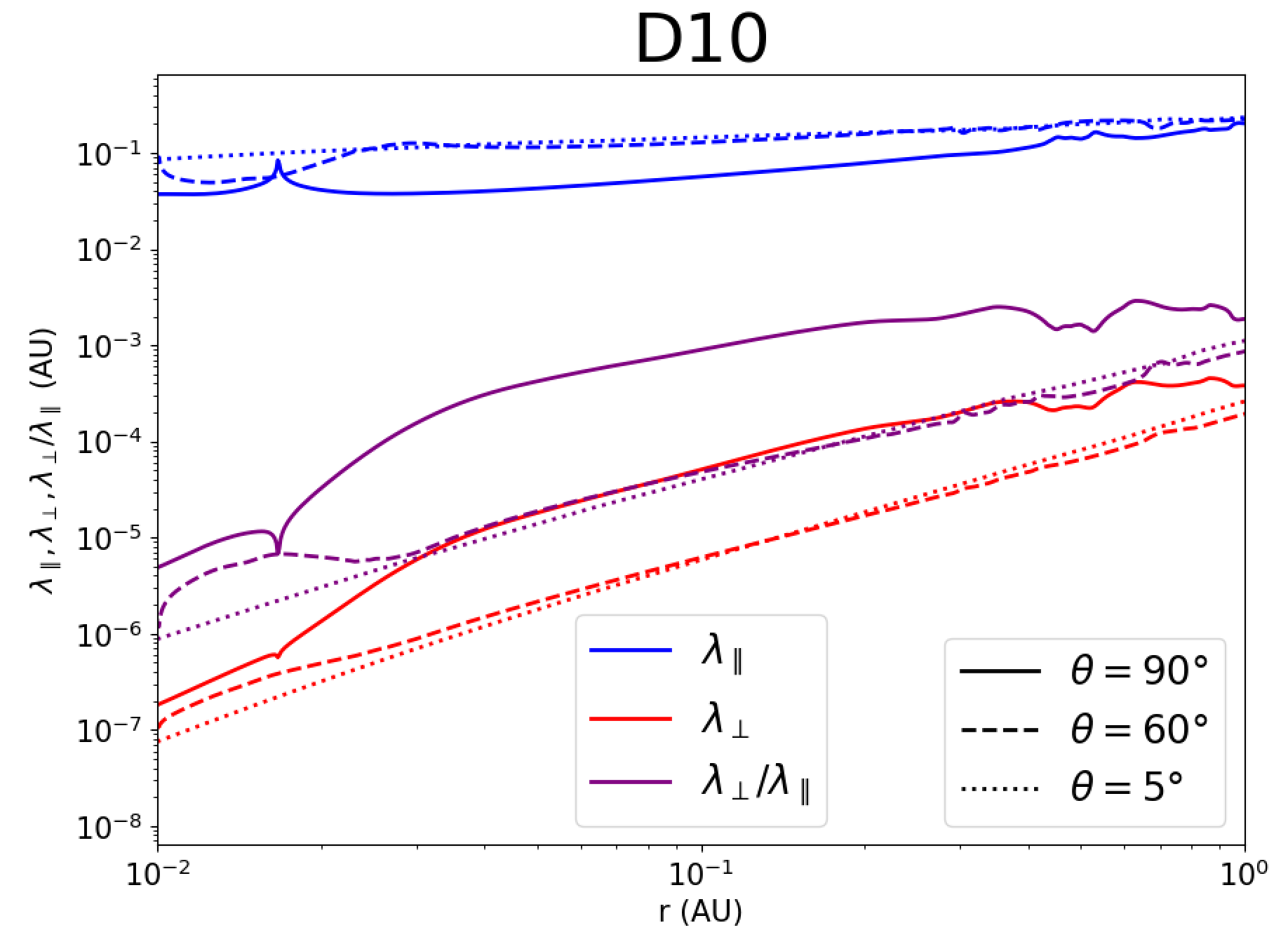}}
    \caption{Radial cuts of the parallel mfp $\lambda_\parallel$ (in blue), the perpendicular mfp $\lambda_\perp$ (in red) and the ratio between them $\lambda_\perp/\lambda_\parallel$ (in purple) in AU for a proton of rigidity 445 MV (which means 100 MeV energy). The solid lines are cuts at the equator ($\theta=90\degree$), the dashed lines are cuts at $\theta=60\degree$ and dotted lines are cuts near the poles ($\theta=4\degree$). The panel on the left correspond to the case D1. The panel on the right correspond to the case D10.}
    \label{fig:ampli_1d}
\end{figure}

Figure \ref{fig:ampli_1d} shows radial cuts of $\lambda_\parallel$ in blue, $\lambda_\perp$ in red and the ratio $\lambda_\perp/\lambda_\parallel$ in purple at three different co-latitudes ($\theta=90\degree$ in solid line, $\theta=60\degree$ in dashed line and $\theta=4\degree$ in dotted line). We see more clearly that the amplitude of the two mfps has changed. With $B_*$ 10 times stronger, $\lambda_\parallel$ has increased by 60\% at all latitudes while $\lambda_\perp$ loses 1 order of magnitude. However the general radial trend far from the Sun does not change with the amplitude, especially at $\theta=60\degree$ and $\theta=4\degree$ : at 1 AU, the parallel mfp evolves as $\propto r^{0.15}$ and the perpendicular mfp evolves as $\propto r^{1.8}$. Here also we can clearly see the effect of the current sheet spread. For case D1, the only latitude at which we see a variation in the trend is the equator (in solid line). We see a bump at $r=2\times10^{-2}$ AU ($3\times10^6$ km) where $\lambda_\parallel$ decreases by 30\% and where $\lambda_\perp$ increases of almost 2 orders of magnitude ; the width of the bump is of $1.5\times10^{-2}$ AU ($2.25\times10^6$ km). There is also a slight increase of $\lambda_\parallel$ by 30\% right before the bump, as well as a slight decrease of 10\% of $\lambda_\perp$. For case D10, the equator is affected, but also the latitude $\theta=60\degree$ because of the edges of the current sheet which is now much more extended. Here $\lambda_\parallel$ has no decrease bump, only a slight increase at $r=1.75\times10^{-2}$ AU, while we see an increase of 2 orders of magnitude for $\lambda_\perp$ from $10^{-7}$ AU to $10^{-5}$ AU (15 km to $1.5\times10^3$ km). Moreover there is no sudden bump, the variation starts at $r=2\times10^{-2}$ AU ($3\times10^6$ km) and is visible until $r=0.9$ AU ($1.35\times10^8$ km). To sum up, the change in amplitude has an impact on the amplitude of the mfps and on their radial distribution in the current sheet and on its edges.

The ratio $\lambda_\perp/\lambda_\parallel$ (purple line) is never greater than 1 in our simulations : for case D1, it is between $10^{-5}$ and $10^{-2}$ ; for case D10, it is between $10^{-6}$ and $10^{-3}$. It is the general behavior expected, except for some specific regions \citep{dwyer_perpendicular_1997, zhang_perpendicular_2003} which could be local structures involved in the diminishing of SEPs \citep{zhang_propagation_2009}. However this simulation does not seem to have that kind of structure, contrary to \cite{chhiber_cosmic-ray_2017} where they had a strong dipole inclined at $30\degree$ ; this kind of structures were visible when crossing the current sheet. This may be due to our modeling of the turbulence, which is not self-consistent with the wind model and may not be realistic enough to describe properly the current sheet.  

\subsection{Influence of geometry}
\label{subsec:study_topo}

Now we want to focus on the influence of geometry, as the solar cycle goes from mainly dipolar at minimum to mostly quadrupolar at maximum \citep{derosa_solar_2012}. To do so, we will focus on cases D1 and Q1 described in table \ref{tab:study_cases}. We still focus on protons of rigidity 445 MV, which corresponds to 100 MeV energy.

\begin{figure}[!t]
    \centering
    \subfigure{\includegraphics[width=0.45\columnwidth]{2d_dip_05.png}}
    \subfigure{\includegraphics[width=0.45\columnwidth]{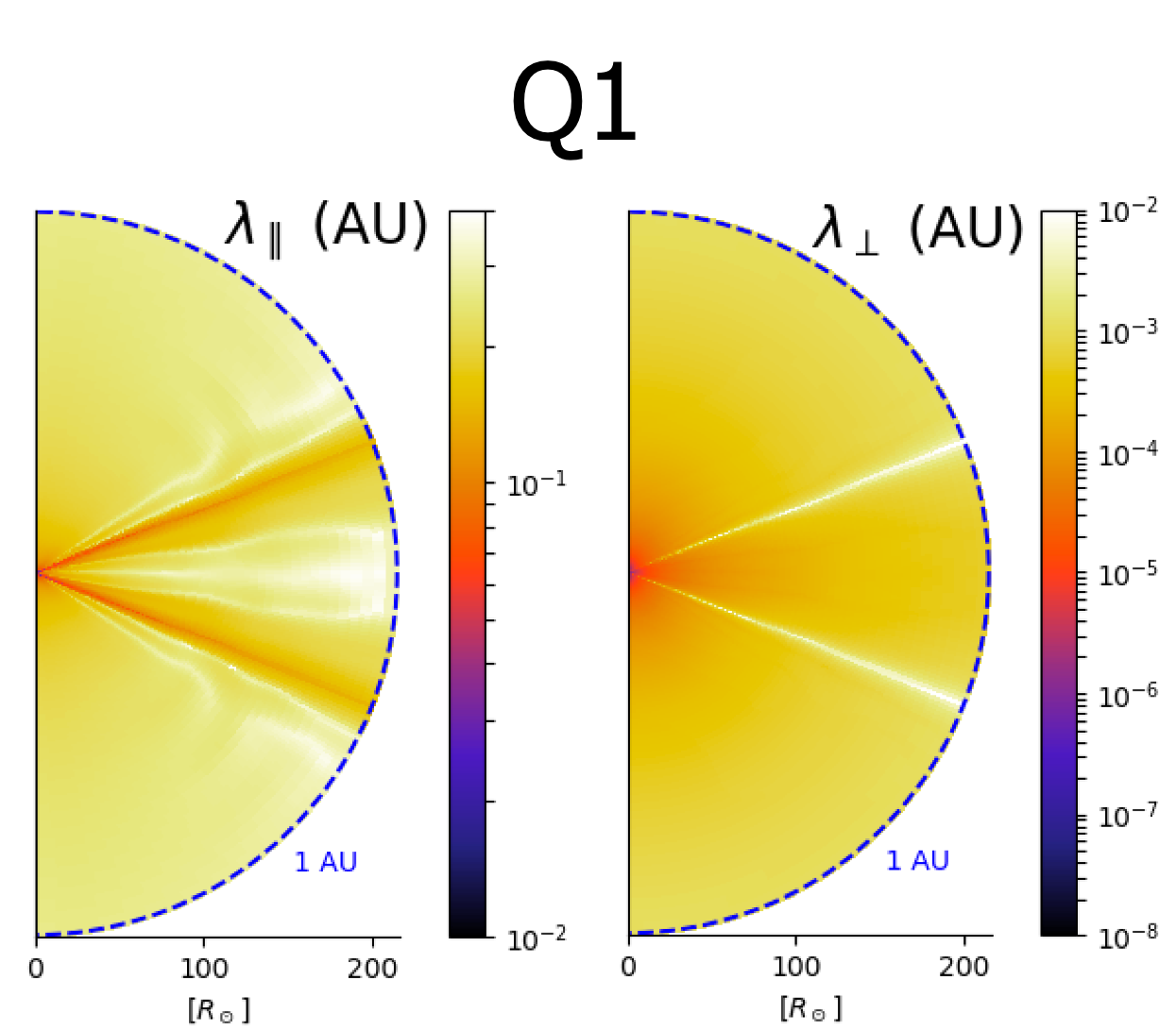}}
    \caption{Meridional cuts of the parallel mfp $\lambda_\parallel$ and the perpendicular mfp $\lambda_\perp$ in AU for a proton of rigidity 445 MV (which means 100 MeV energy). The two panels on the left correspond to the case D1, the two panels on the left correspond to the case Q1.}
    \label{fig:topo_2d}
\end{figure}

Figure \ref{fig:topo_2d} shows the 2D meridional cuts of $\lambda_\parallel$ and $\lambda_\perp$ in units of AU for case D1 on the left and case Q1 on the right. What is immediately striking is that the position of the current sheet is different : for case D1, there is only one current sheet at the equator, while for case Q1 there are two current sheets at $\theta=60\degree$ and $\theta=120\degree$. There is thus a new zone delimited by the two current sheets where the diffusion is different compared to the high latitude regions : $\lambda_\parallel$ is higher and $\lambda_\perp$ is slightly lower. We will quantify this in more details with the next figure.

\begin{figure}[!t]
    \centering
    \subfigure{\includegraphics[width=0.45\columnwidth]{sum_d1.png}}
    \subfigure{\includegraphics[width=0.45\columnwidth]{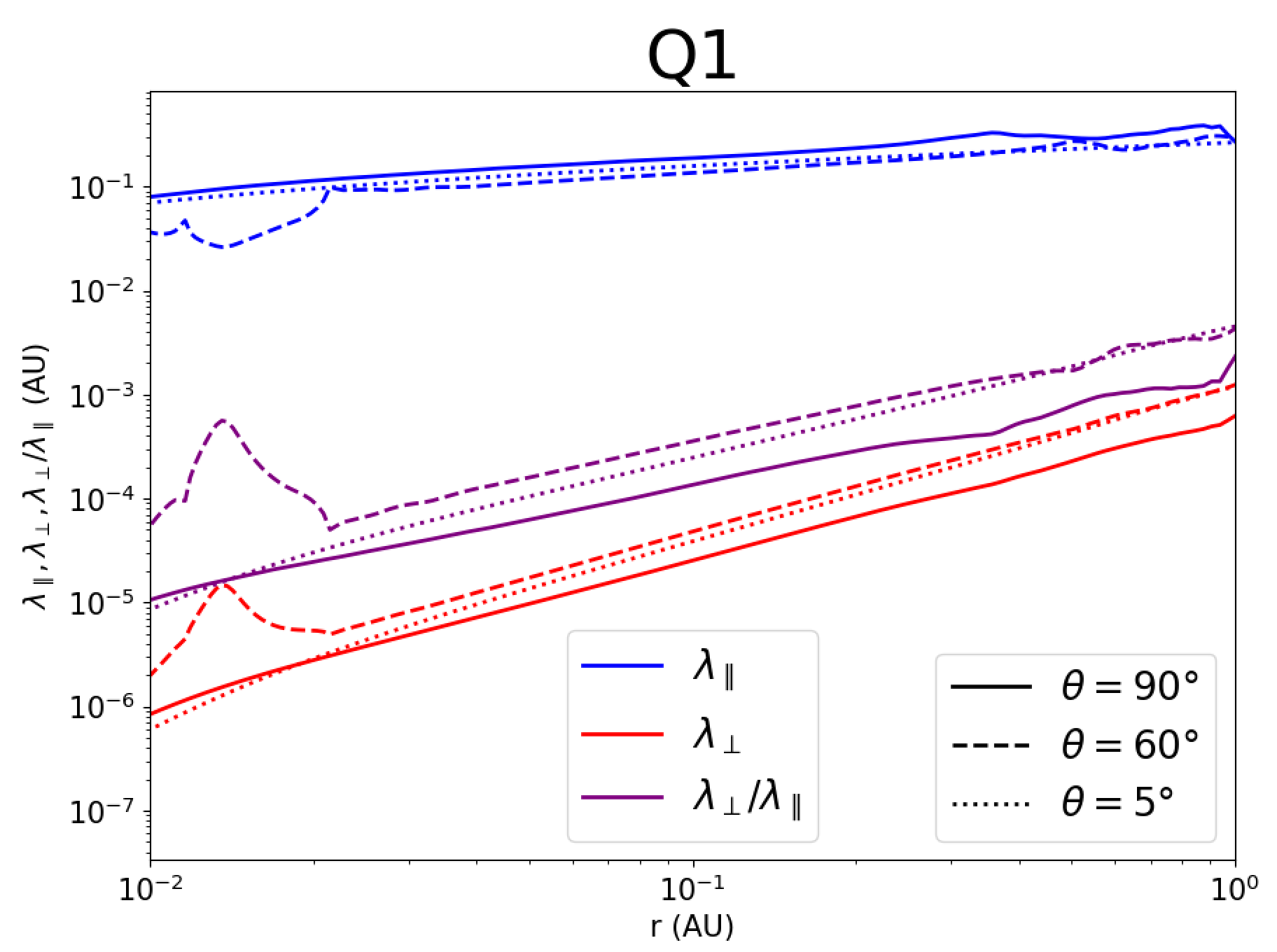}}
    \caption{Radial cuts of the parallel mfp $\lambda_\parallel$ (in blue), the perpendicular mfp $\lambda_\perp$ (in red) and the ratio between them $\lambda_\perp/\lambda_\parallel$ (in purple) in AU for a proton of rigidity 445 MV (which means 100 MeV energy). The solid lines are cuts at the equator ($\theta=90\degree$), the dashed lines are cuts at $\theta=60\degree$ and dotted lines are cuts near the poles ($\theta=4\degree$). The panel on the left correspond to the case D1. The panel on the right correspond to the case Q1.}
    \label{fig:topo_1d}
\end{figure}

Figure \ref{fig:topo_1d} shows radial cuts of $\lambda_\parallel$ in blue, $\lambda_\perp$ in red and the ratio $\lambda_\perp/\lambda_\parallel$ in purple at three different co-latitudes ($\theta=90\degree$ in solid line, $\theta=60\degree$ in dashed line and $\theta=4\degree$ in dotted line). We can see that the amplitudes are very similar in both cases : $\lambda_\parallel$ varies between $2\times10^{-2}$ and $2\times10^{-1}$ AU while $\lambda_\perp$ varies between $10^{-6}$ and $10^{-2}$ AU. However for case Q1 $\lambda_\perp$ reaches only $10^{-3}$: we can see with figure \ref{fig:topo_2d} that it is less enhanced in the current sheets than in case D1. The bump is also slightly closer to the Sun at $r=1.5\times10^{-2}$ AU. The latitudinal distribution is very different due to the disposition of the current sheets : the variations observed at $\theta=90\degree$ are now seen at $\theta=60\degree$. The equatorial plane in case Q1 is more similar to the polar region due to the fact that it is now between two current sheets. The radial trend remains the same as described before. Hence the change in geometry is affecting mostly the latitudinal distribution of CRs diffusion.

So in the end changing amplitude versus changing geometry are not likely to be equivalent. It means that there is some hope to combine our simulations to CR transport models to correlate past CR records on Earth with past evolution of $B_\odot$ in terms of amplitude and geometry.

\subsection{Influence of energy}
\label{subsec:study_energy}

We will now vary the rigidity/energy of the protons to see how SEPs and GCRs might be affected differently. 

\begin{figure}[!t]
    \centering
    \subfigure{\includegraphics[width=0.45\columnwidth]{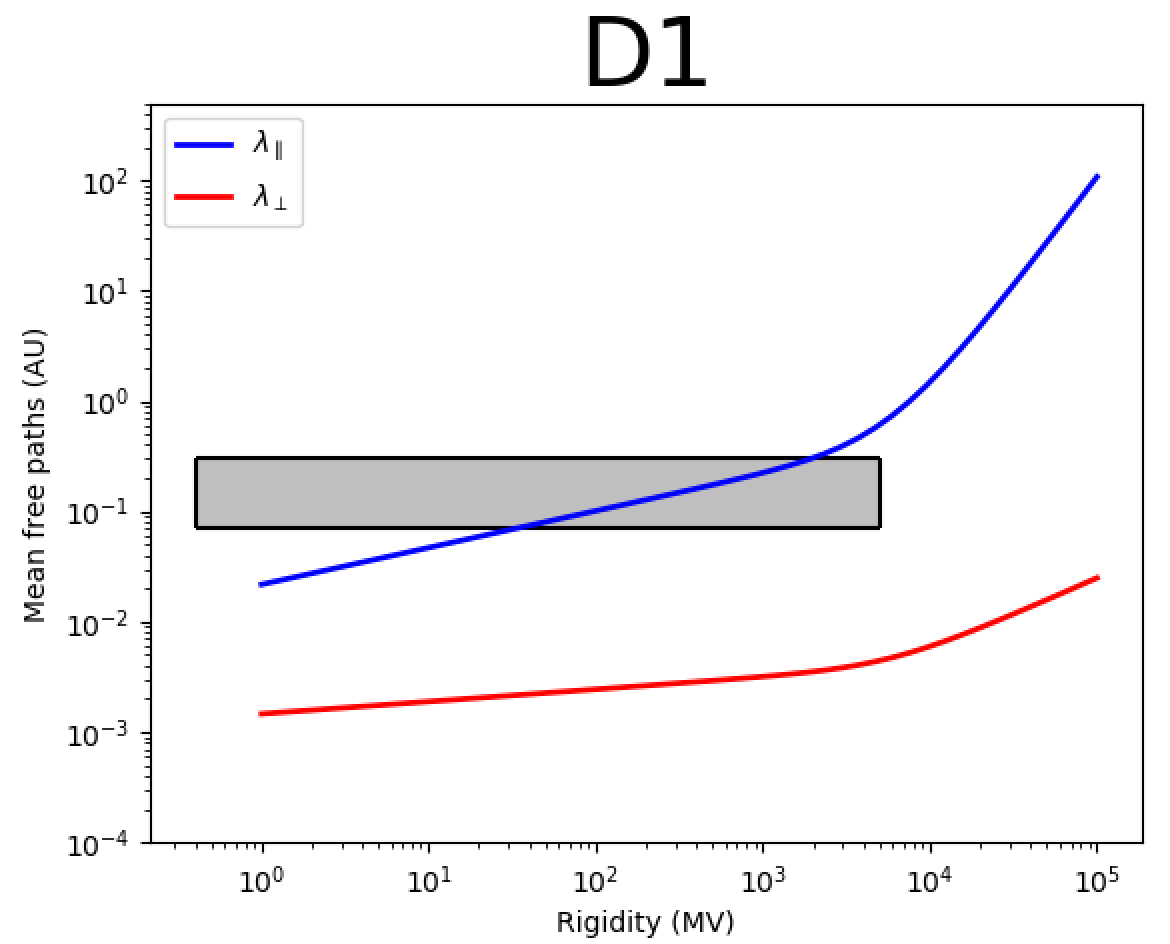}}
    \subfigure{\includegraphics[width=0.45\columnwidth]{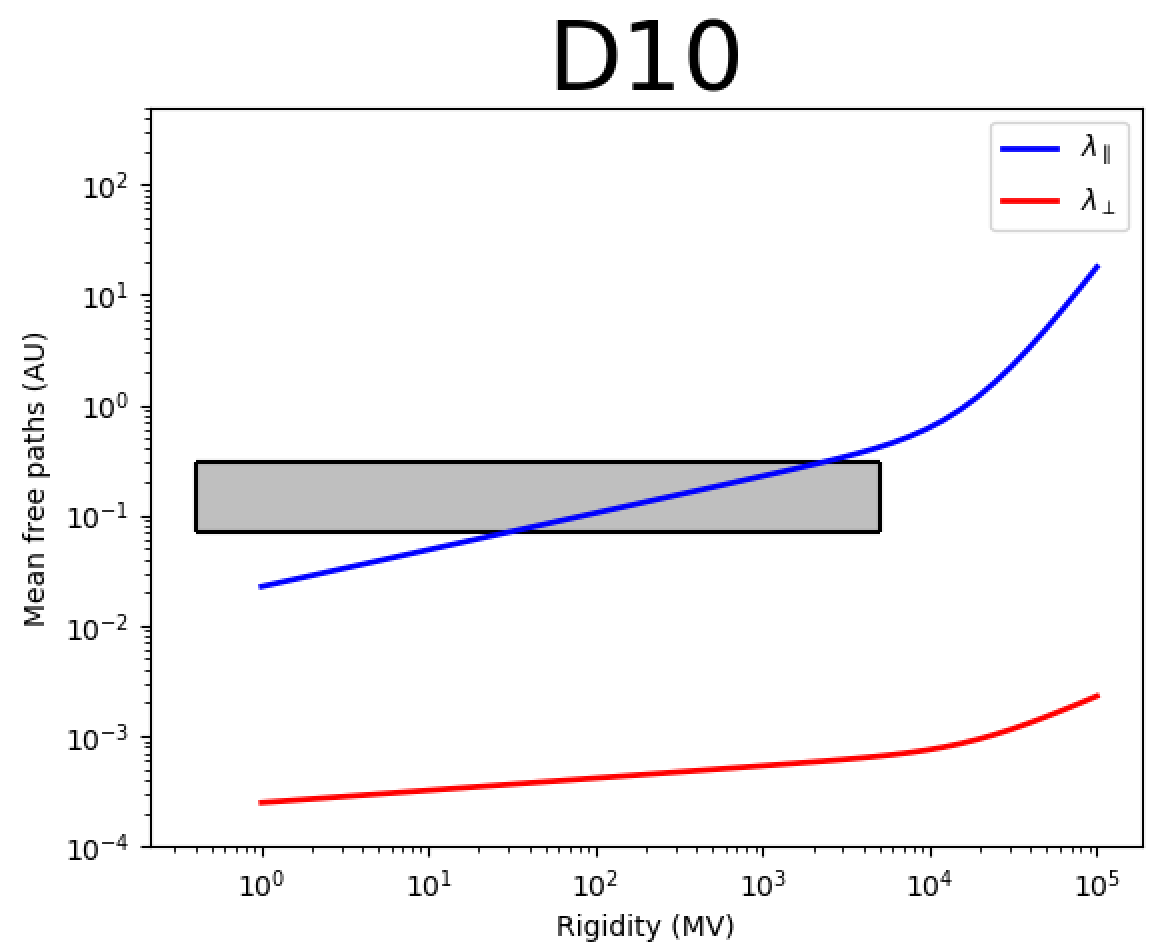}}
    \caption{Rigidity dependency of the parallel mfp $\lambda_\parallel$ (in blue) and the perpendicular mfp $\lambda_\perp$ (in red) in the equatorial plane at 1 AU for the case D1 (on the left) and case D10 (on the right). The gray area corresponds to the Palmer consensus for reference to observational data for the parallel mean free path \citep{palmer_transport_1982}. 
    }
    \label{fig:rig_1d}
\end{figure}

Figure \ref{fig:rig_1d} shows the rigidity dependency of the two mfps between 1 and $10^5$ MV, which corresponds to between 1 keV and 100 GeV. They are plotted at 1 AU in the equatorial plane, on the left for case D1 and on the right for case D10. For reference, we have also plotted the Palmer consensus range \citep{palmer_transport_1982} as a grey rectangle, which corresponds to the range observed in data for the parallel mean free path only. For case D1, $\lambda_\parallel$ goes from $2\times10^{-2}$ AU at 1 MV to $200$ AU at $10^5$ MV, first with a relation of $\propto P^{0.33}$, then at $2\times10^3$ MV breaks into the relation $\propto P^{1.31}$. $\lambda_\perp$ has a weaker sensitivity to rigidity, going from $1\times10^{-3}$ AU at 1 MV to $2\times10^{-2}$ AU at $10^5$ MV, starting with a relation of $\propto P^{0.33}$, then breaking at also $2\times10^3$ MV to go with the relation $\propto P^{1.36}$. The break at $2\times10^3$ MV corresponds to the transition between SEPs and GCRs at 1 GeV \citep{zank_radial_1998}. For case D10, the values of $\lambda_\parallel$ are smaller, going from 0.02 Au to 20 AU, and the values of $\lambda_\perp$ are also smaller, going from $3\times10^{-4}$ AU to $2\times10^{-3}$ AU. The break also happens at a different rigidity, here around $2\times10^4$ MV. The case D1 is closer to the simulations of \cite{bieber_nonlinear_2004} : $\lambda_\parallel$ goes from $2\times10^{-1}$ AU at 1 MV to $10$ AU at $9\times10^4$ MV, while $\lambda_\perp$ goes from $2\times10^{-3}$ AU at 1 MV to $1\times10^{-2}$ AU at $10^5$ MV. Case D10 on the other hand is closer to the results of \cite{chhiber_cosmic-ray_2017}: $\lambda_\parallel$ goes from $0.05$ AU at 1 MV to $20$ AU at $10^4$ MV, while $\lambda_\perp$ goes from $2\times10^{-3}$ AU at 1 MV to $5\times10^{-3}$ AU at $10^5$ MV. We see more difference with $\lambda_\perp$ because we chose a different modeling from theirs. This shows that the energy dependency is very sensitive to the intensity of the magnetic field, especially at 1 AU. 

\begin{figure}[!t]
    \centering
    \subfigure{\includegraphics[width=0.45\textwidth]{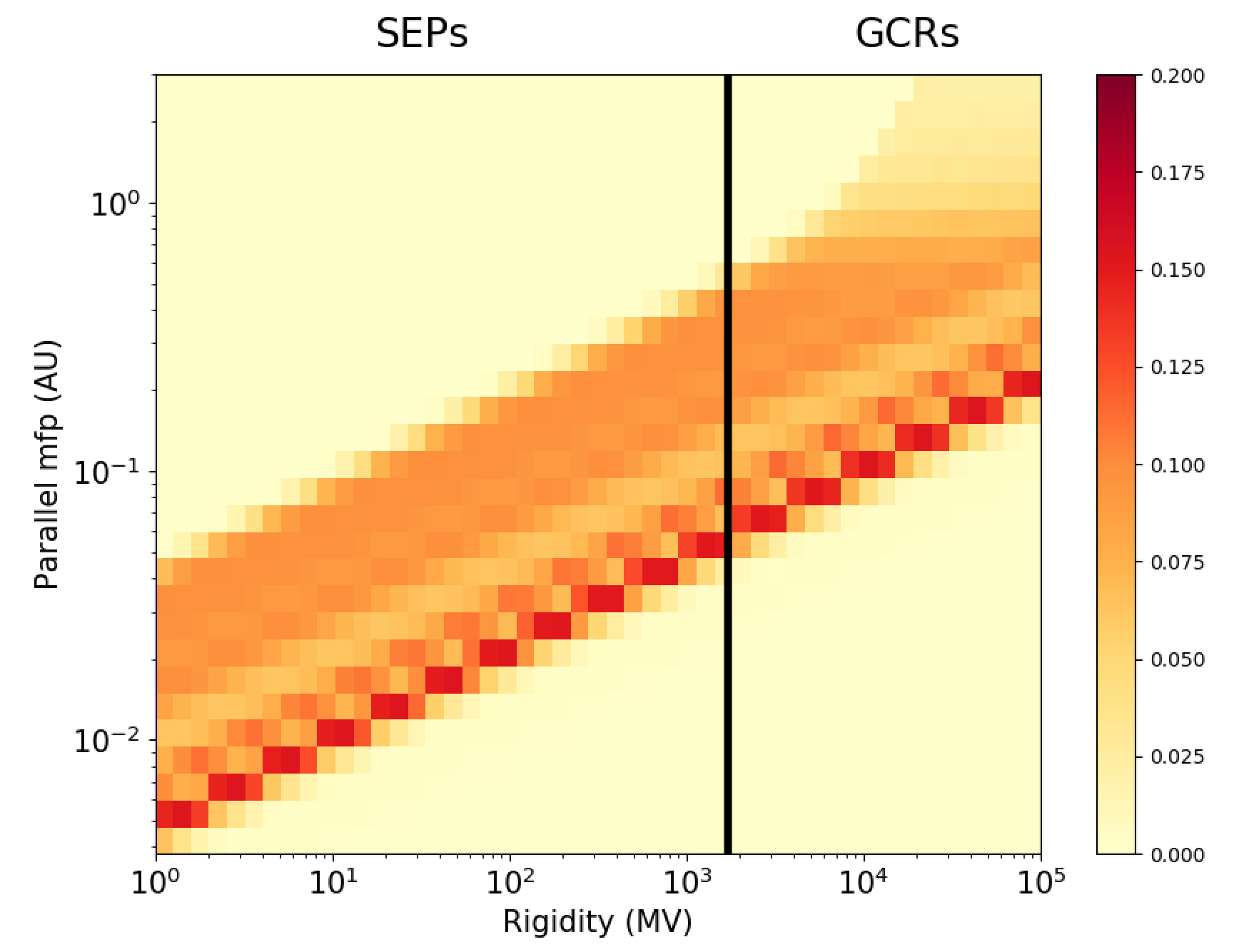}}
    \subfigure{\includegraphics[width=0.48\textwidth]{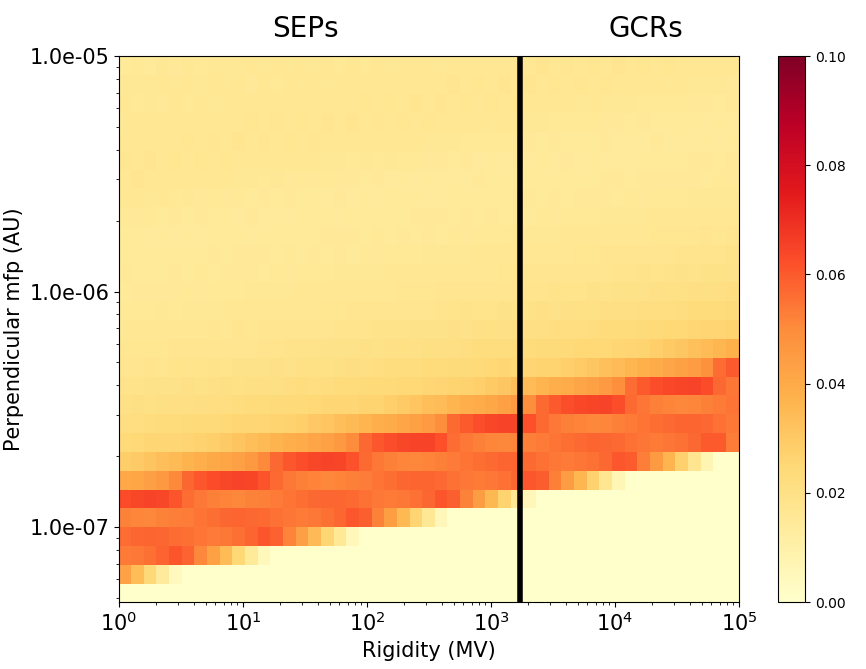}}
    \caption{Collection of 1D histograms of the distribution of the parallel mfp $\lambda_\parallel$ (on the left) and the perpendicular mfp $\lambda_\perp$ (on the right) depending on rigidity for the case D1. SEPs and GCRs are separated using a vertical black line at $P=1.69\times10^3$ MV.}
    \label{fig:rig_2d}
\end{figure}

Figure \ref{fig:rig_2d} shows a collection of 1D histograms of the distribution of the values of $\lambda_\parallel$ (on the left) and $\lambda_\perp$ (on the right) depending on the rigidity of the proton considered. We consider once again particles with rigidities between 1 and $10^5$ MV, which corresponds to energies between 1 keV and 100 GeV. Thus, on these histograms, SEPs are located on the left on GCRs on the right, separated by a vertical black line at $P=1.69\times10^3$ MV (which corresponds to a 1 GeV energy). These histograms are made considering all the values of the mfps in the azimuthally-averaged meridional plane between 1 and 220 solar radii and from one pole to the other. We see for the parallel mfp that the distribution of values for SEPs is more centered around small values: around 20\% of the distribution lies between 0.004 and 0.05 AU for rigidities under $10^3$ MV, and between 10 and 15\% of the distribution lies within slightly bigger values between 0.01 and 0.8 AU. For GCRs, the trend is the same, with 20\% of the distribution between 0.05 and 0.2 AU, and between 10 and 15\% of the distribution between 0.1 and 1 AU. However, the more energetic is the proton, the more spread is the distribution of values, with a drift of the tail of the distribution. For the perpendicular mfp, we can see the weaker dependency on rigidity, as SEPs and GCRs display almost the same behavior : for SEPs, 8\% of the distribution is between $1\times10^{-7}$ and $3.0\times10^{-7}$ AU; for GCRs, 8\% of the distribution is between $3.0\times10^{-7}$ AU and $5.0\times10^{-7}$ AU. Compared to $\lambda_\parallel$, the distribution is much more spread towards higher values of mfps. Our models thus allows us to probe the entire heliosphere to derive statistical information about both SEPs and GCRs, which will come useful when computing a realistic count of cosmic rays reaching the Earth.

\section{Application to real configurations}
\label{sec:config}

\begin{figure}[!t]
    \centering
    \subfigure{\includegraphics[width=0.45\textwidth]{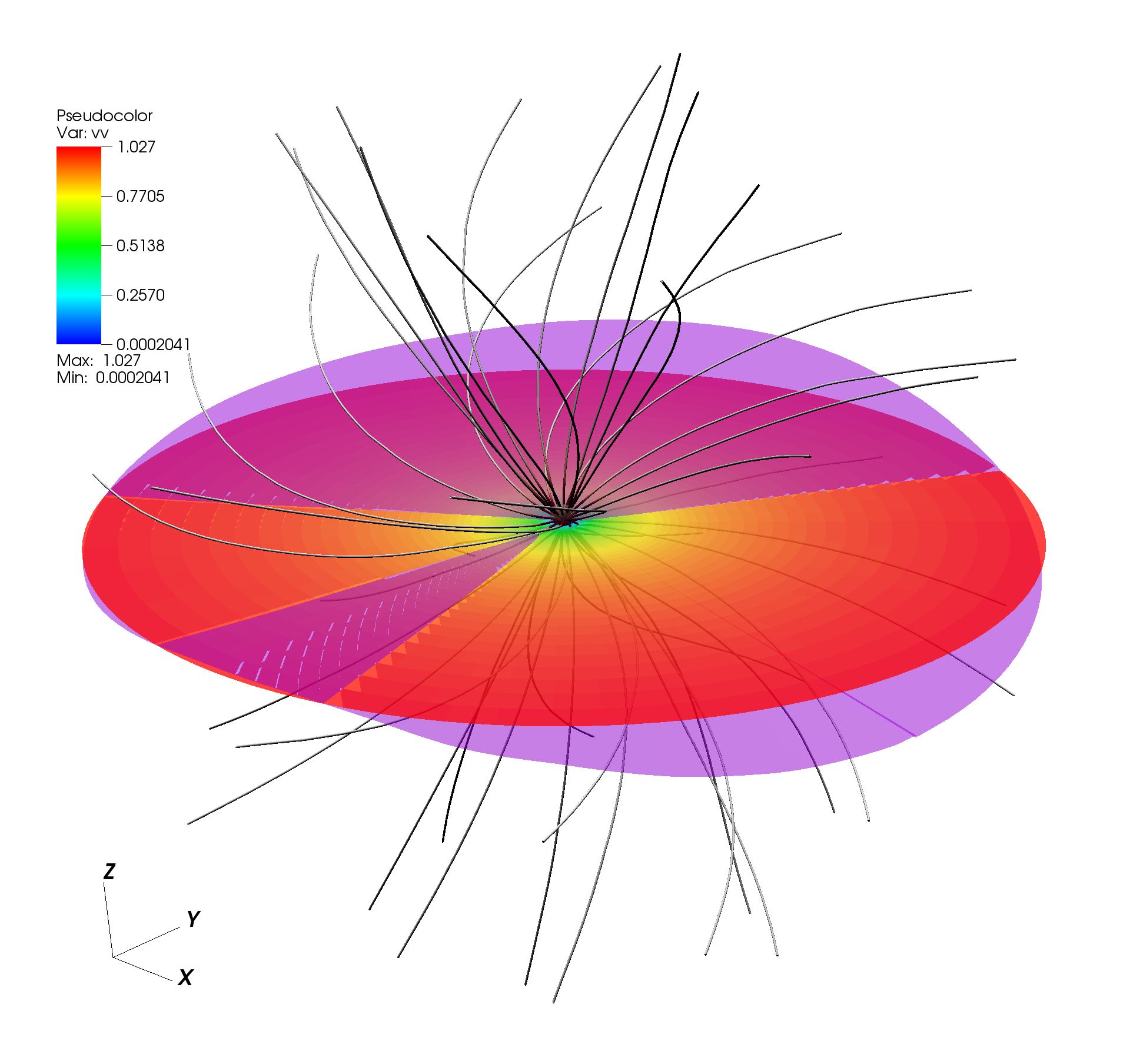}}
    \subfigure{\includegraphics[width=0.45\textwidth]{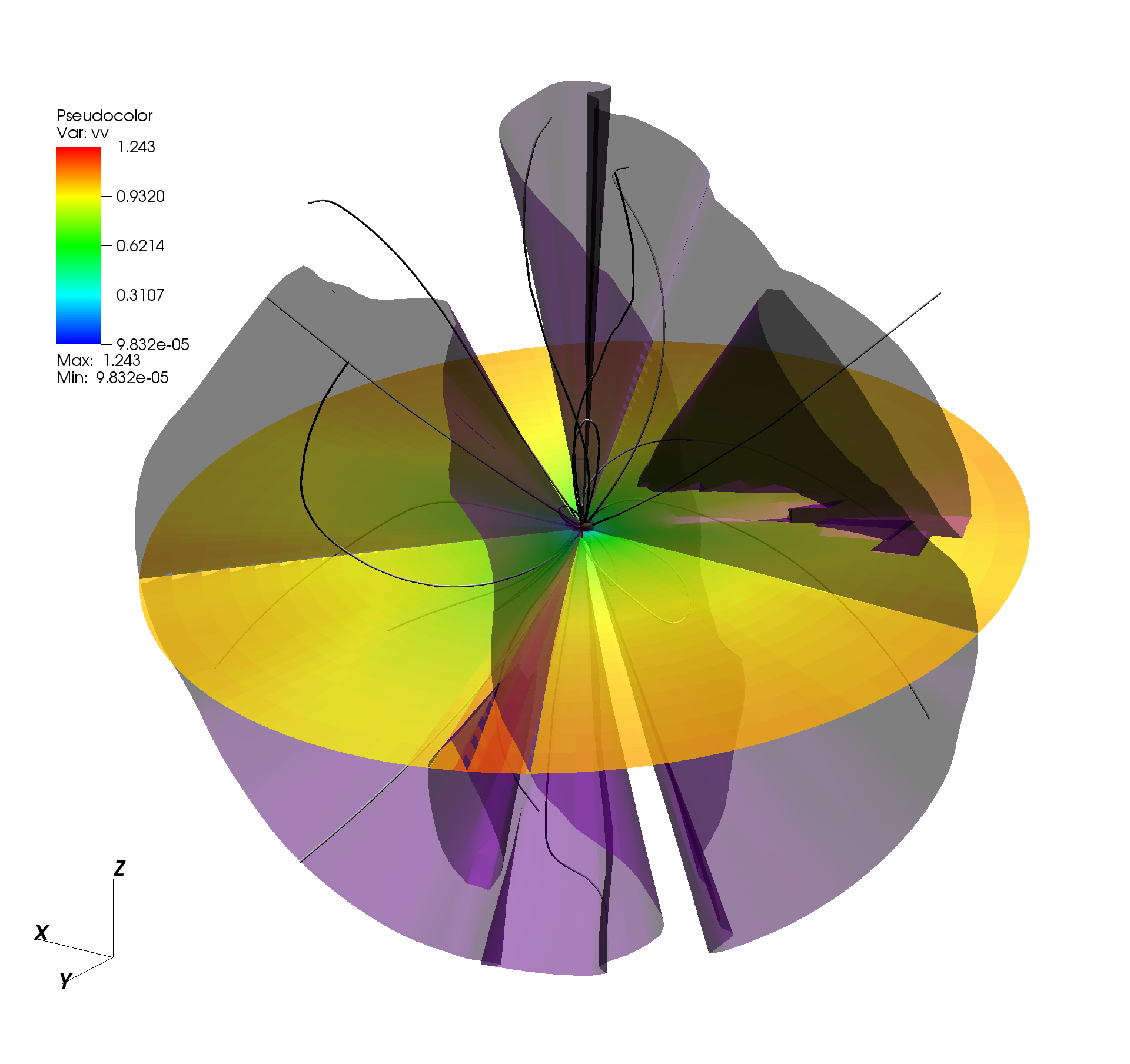}}
    \caption{3D visualizations of the wind simulations performed with synoptic maps from Wilcox observatory. On the left it is the minimum of activity of October 1995, on the right the maximum of activity of August 1999. The color scale corresponds to the wind speed in the ecliptic plane. The magnetic field lines are represented as tubes. The purple surface is the contour of the neutral current sheet.}
    \label{fig:config_3d}
\end{figure}

We will now use synoptic maps from the Wilcox Observatory to see the 3D configuration of CR diffusion for real magnetic field configurations \citep{hoeksema_evolution_2009}. We chose a map corresponding to the minimum of activity reached in October 1995 (with a SSN of 25, 10 erg of energy in the dipole and 0.1 erg of energy in the quadrupole), and a map corresponding to the maximum of activity reached in August 1999 (with a SSN of 100, 3 erg of energy in the dipole and 1 erg of energy in the quadrupole) \citep{derosa_solar_2012}. The corresponding configuration of the wind up to 1 AU can be seen in figure \ref{fig:config_3d}, especially the position of the current sheet. More information about the maps can be found in \cite{reville_global_2017} where robust simulations have already been performed and commented. To use the map in the simulation, we project the magnetic field on spherical harmonics, and then use the decomposition on the first 15 $\ell$ degrees to reconstruct the magnetic field. It is initialized with a potential field source surface (PFSS) method \citep{altschuler_magnetic_1969, schatten_model_1969, schrijver_photospheric_2003} to fill the whole corona, and then the relaxation of the wind modifies the magnetic field according to the MHD equations. These configurations combine the two effects discussed earlier : at minimum of activity, the magnetic field is mostly dipolar with a weaker amplitude ; at maximum of activity, it is mostly quadrupolar with a stronger amplitude. These simulations are also non-axisymmetric, which means that we expect large longitudinal variations. Please note that the fit presented in equation \ref{eq:epsilon_fit} was made based on 2.5D simulations, so without longitudinal variations. To apply this fit to the 3D case, we actually apply the fit to each longitude to reconstruct the 3D approximation of $\epsilon$, which may imply some differences to a fit directly in 3D. 

\begin{figure}[!t]
    \centering
    \subfigure{\includegraphics[width=0.45\columnwidth]{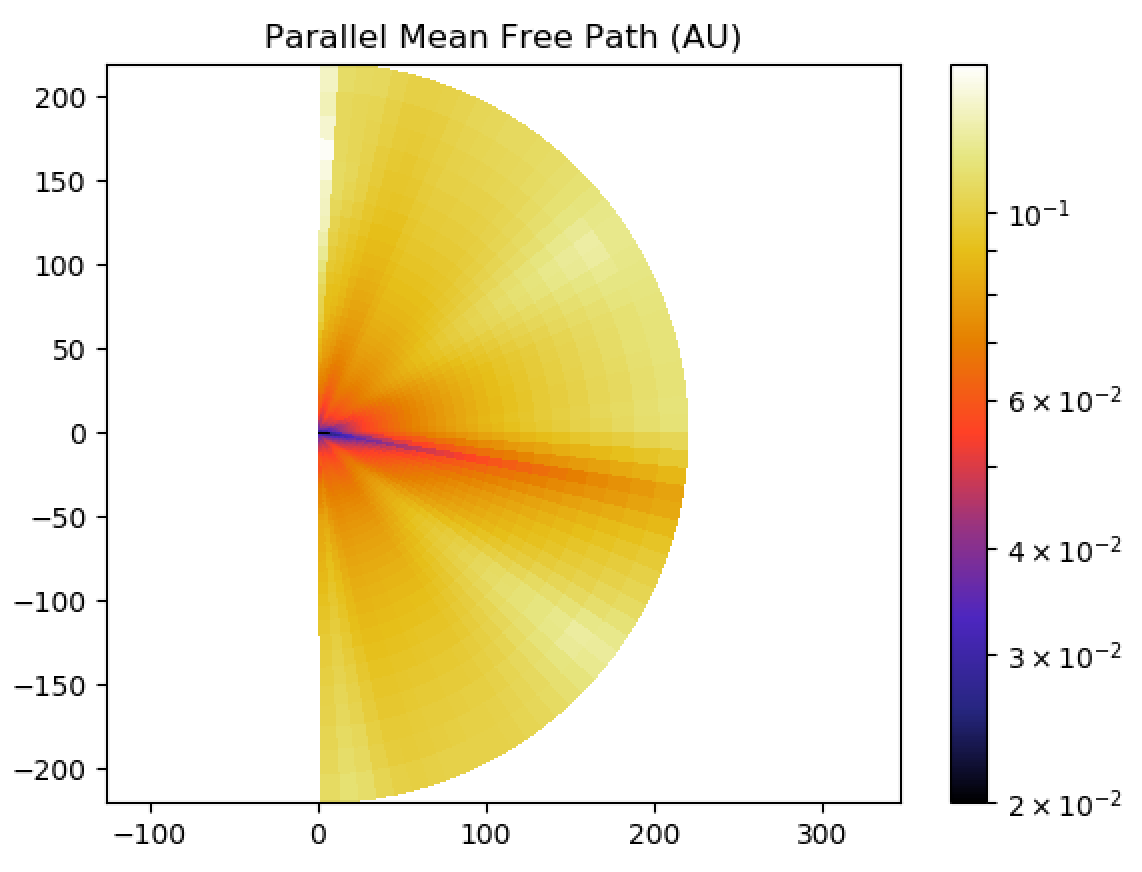}}
    \subfigure{\includegraphics[width=0.45\columnwidth]{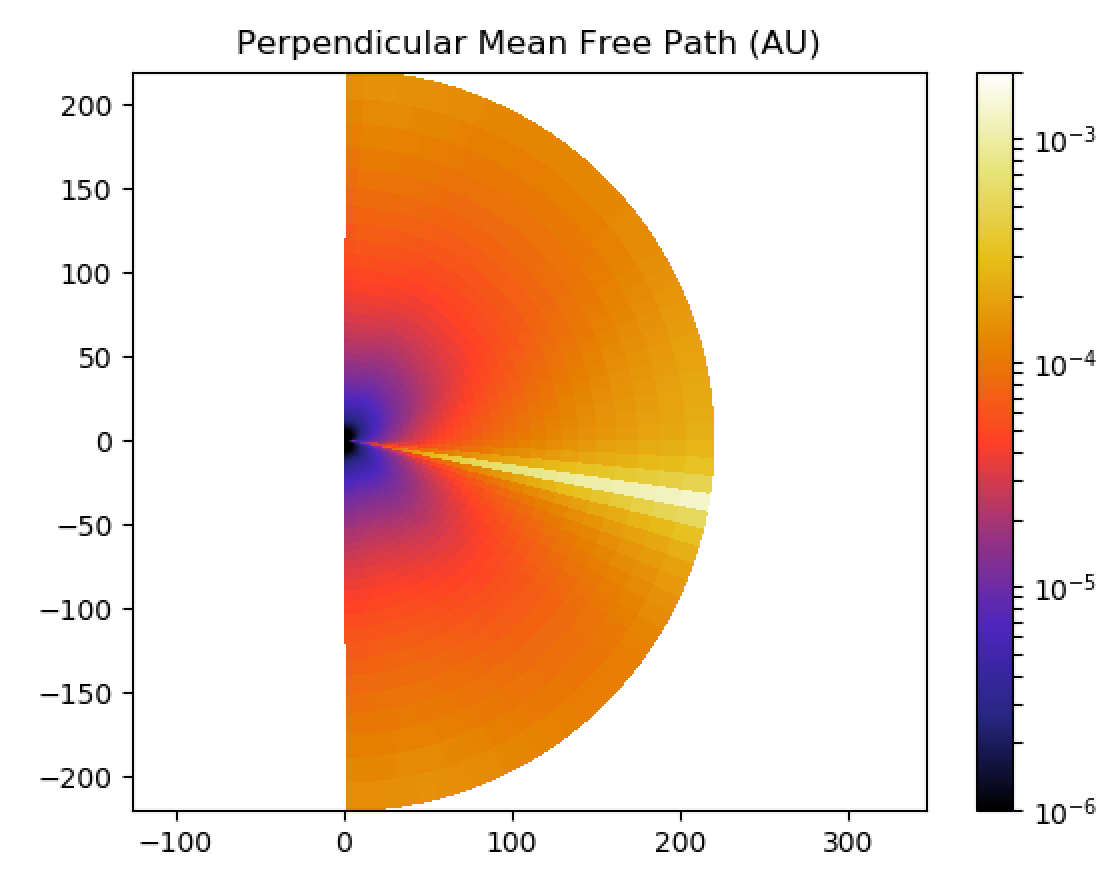}}
    \subfigure{\includegraphics[width=0.45\columnwidth]{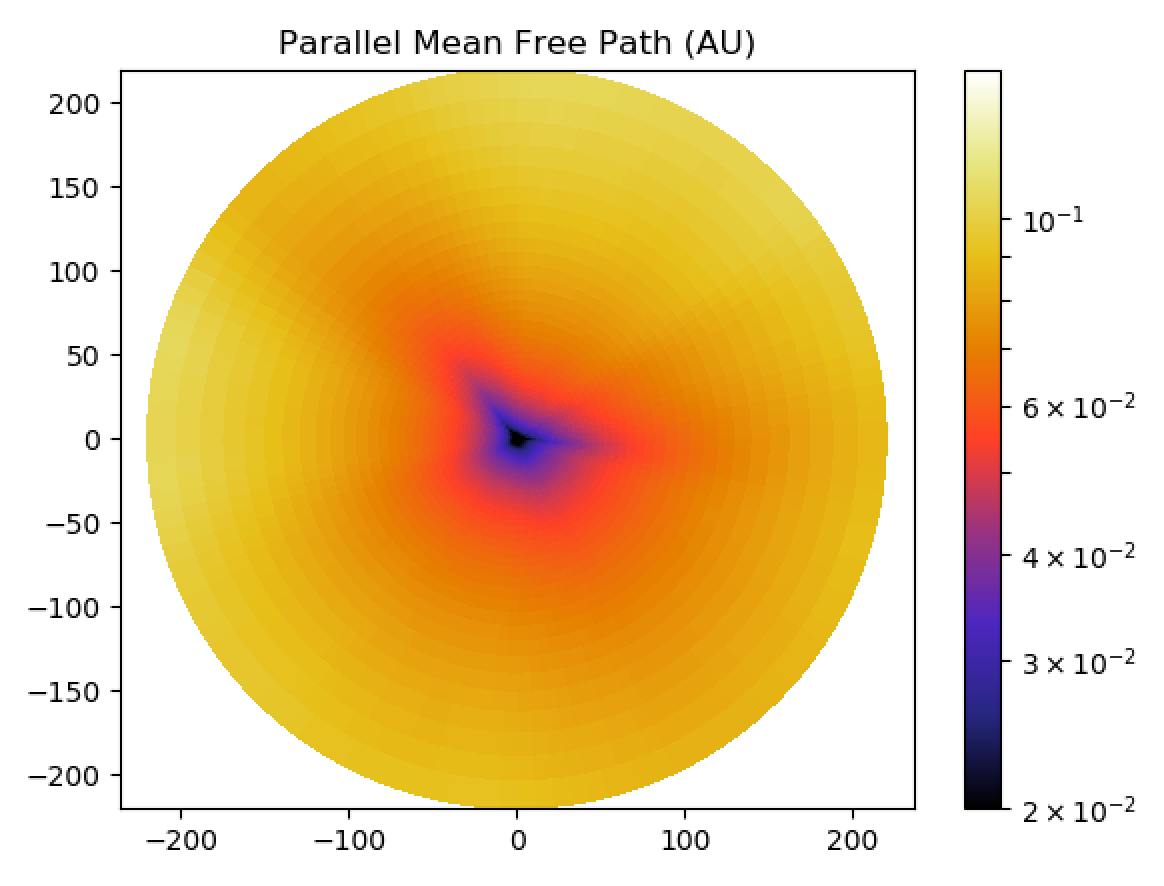}}
    \subfigure{\includegraphics[width=0.45\columnwidth]{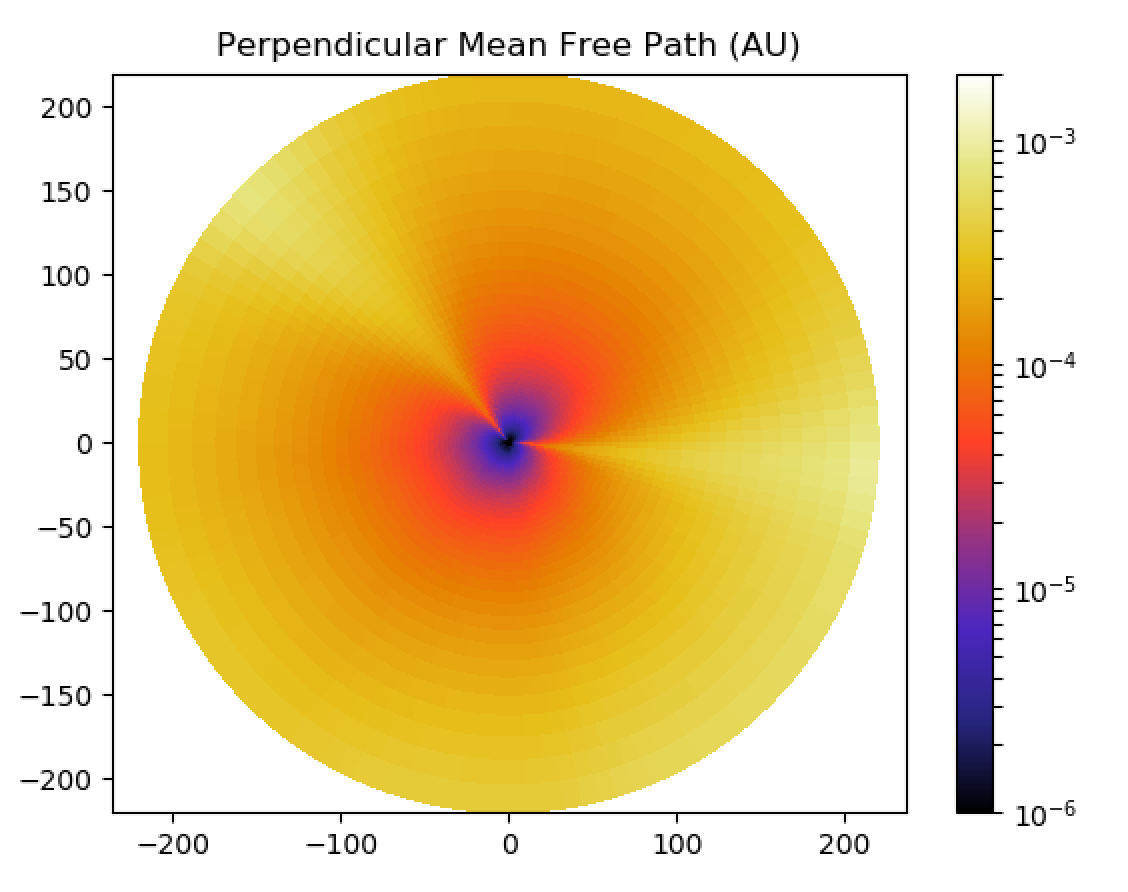}}
    \caption{Meridional and equatorial cuts at respectively $\phi=0$ and $\theta=\pi/2$ of the parallel mfp $\lambda_\parallel$ (on the left) and the perpendicular mfp $\lambda_\perp$ (on the right) in AU for a proton of rigidity 445 MV (which means 100 MeV energy). These panels correspond to a minimum of solar activity reached in October 1995.}
    \label{fig:config_min}
\end{figure}

Figure \ref{fig:config_min} shows the 2D meridional and equatorial cuts at respectively $\phi=0$ and $\theta=\pi/2$ of the parallel (on the left) and perpendicular (on the right) mfps in units of AU for the minimum of activity. The meridional cuts are indeed pretty similar to case D1 (cf. Figure \ref{fig:ampli_2d}, left panel) : we have one current sheet located near the equator which decreases $\lambda_\parallel$ and increases $\lambda_\perp$. The difference is that here the current sheet is inclined, which is expected with a 30-degree shift to the southern hemisphere \citep{mursula_bashful_2003}. The equatorial cut shows that the distribution is rather isotropic, except at $\phi=135\degree$ and $\phi=270\degree$ where we can see local decrease of the parallel mfp. These are due to current sheets as well. This shows that even in minimum of activity, we need to take into account the 3D structure of the magnetic field.

\begin{figure}[!t]
    \centering
    \subfigure{\includegraphics[width=0.45\columnwidth]{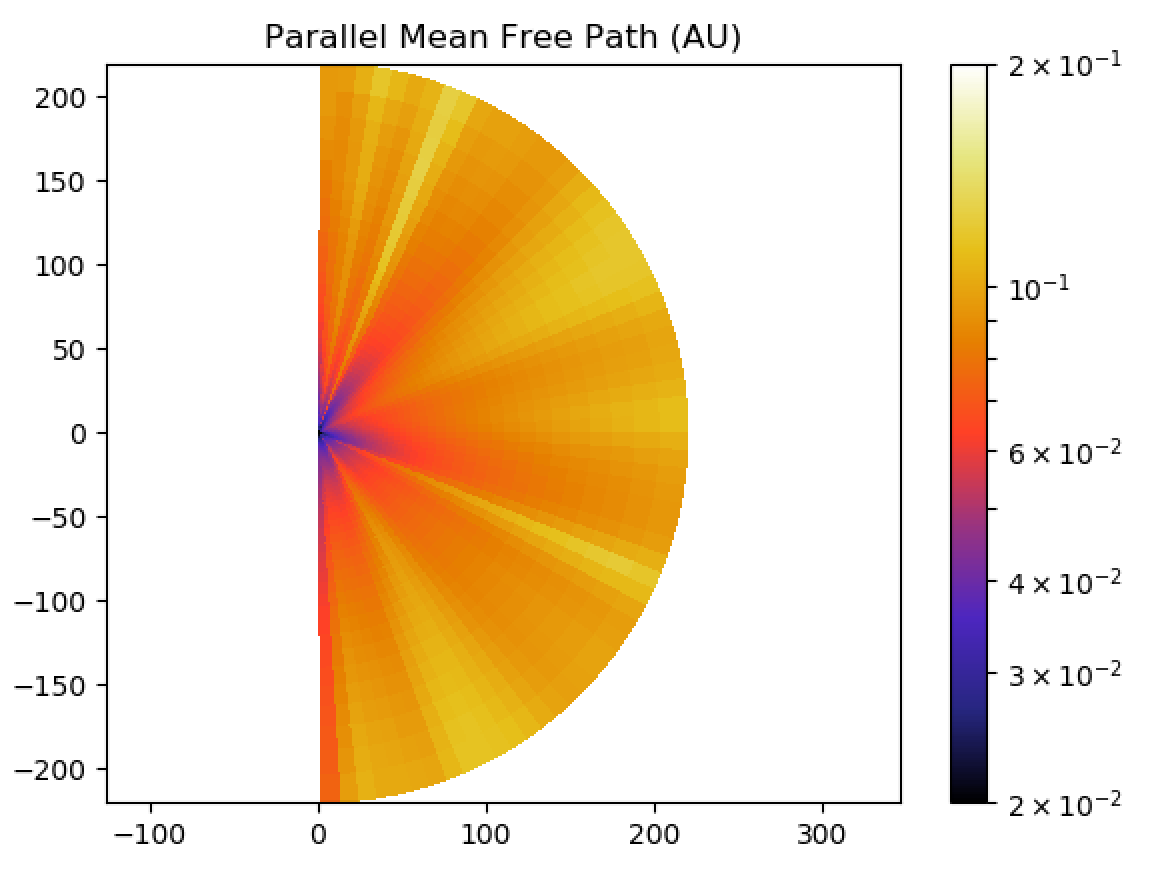}}
    \subfigure{\includegraphics[width=0.45\columnwidth]{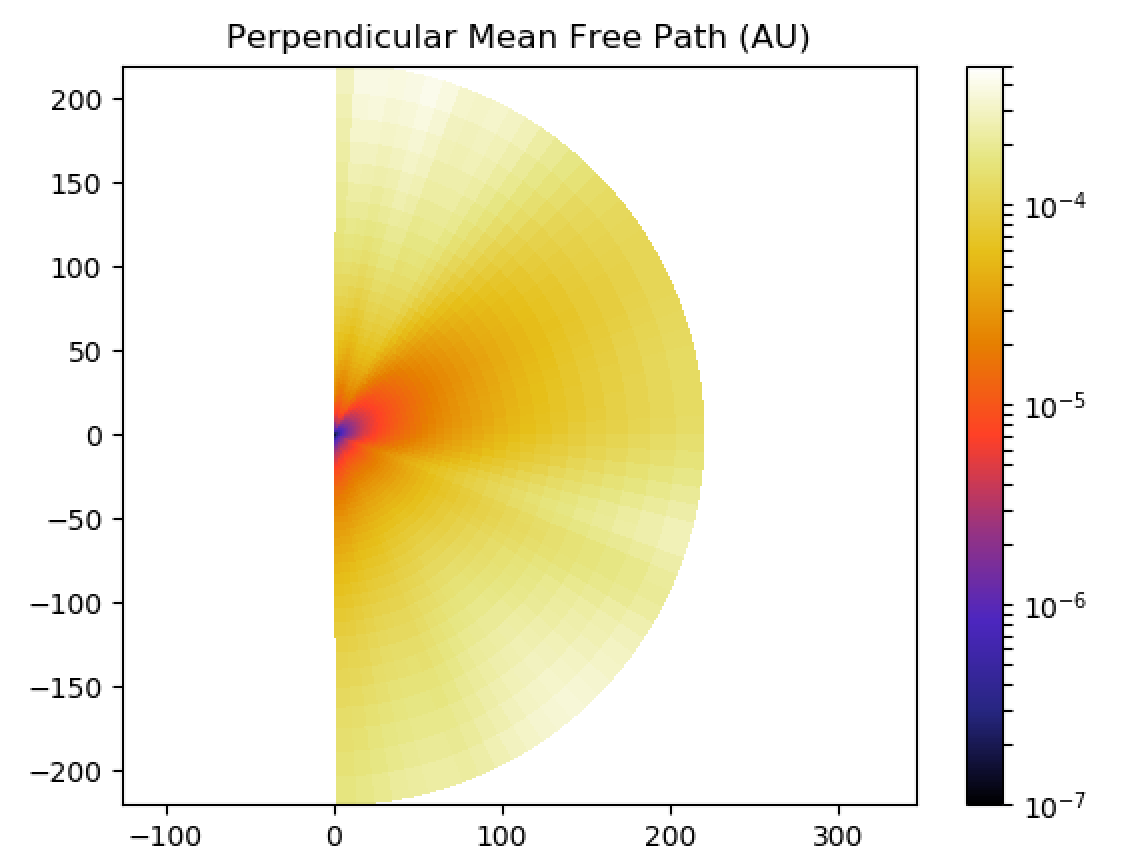}}
    \subfigure{\includegraphics[width=0.45\columnwidth]{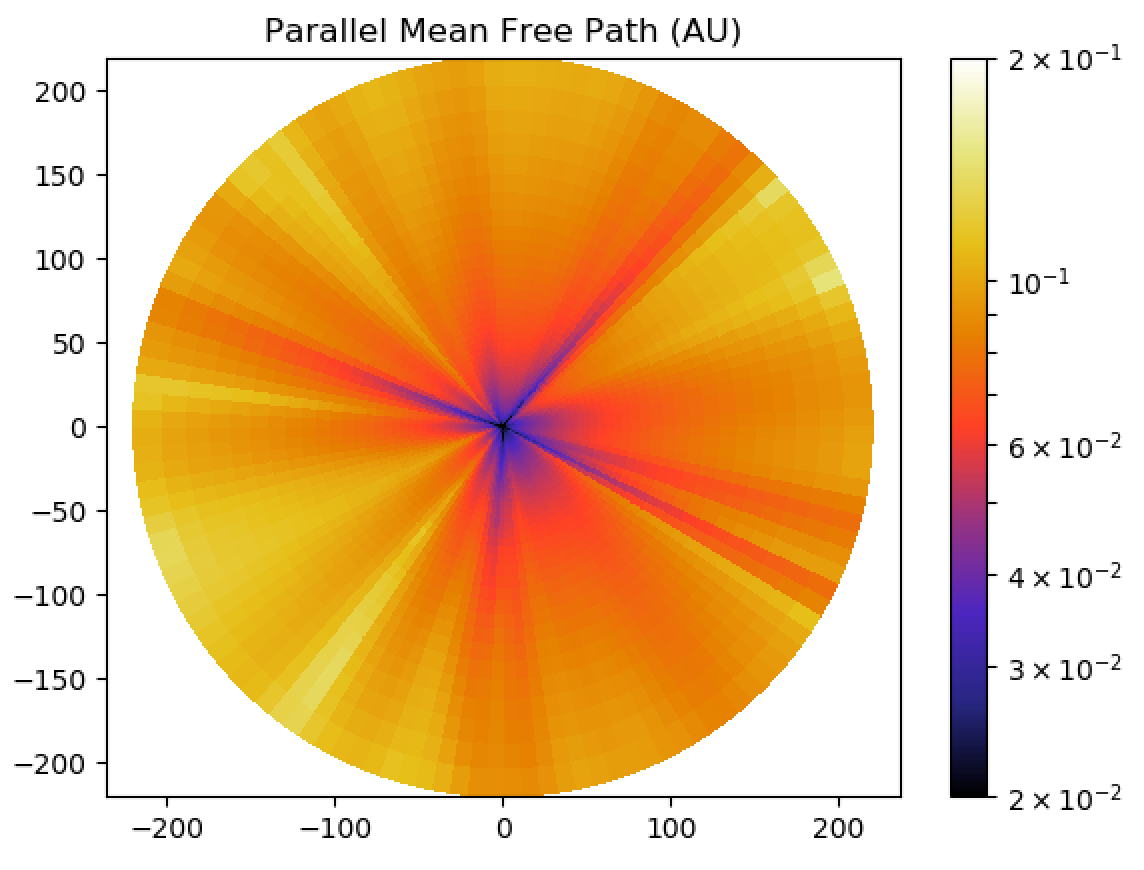}}
    \subfigure{\includegraphics[width=0.45\columnwidth]{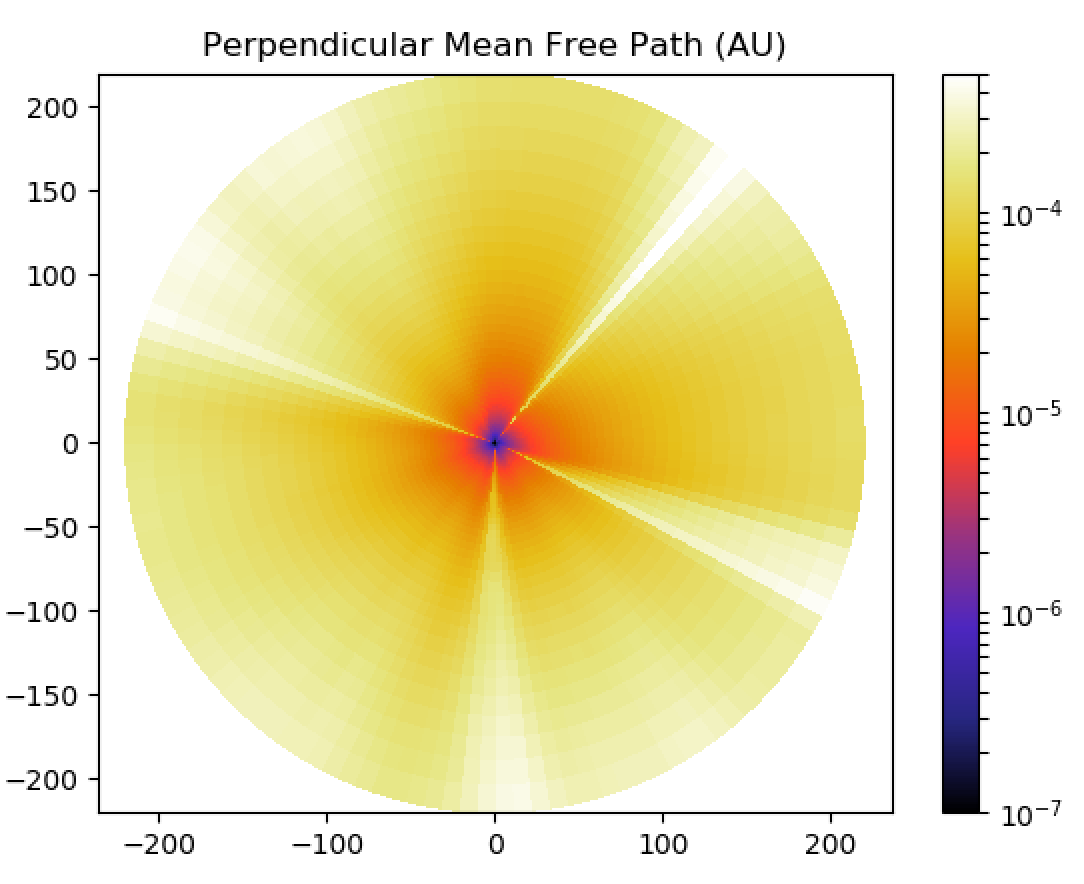}}
    \caption{Meridional and equatorial cuts at respectively $\phi=0$ and $\theta=\pi/2$ of the parallel mfp $\lambda_\parallel$ (on the left) and the perpendicular mfp $\lambda_\perp$ (on the right) in AU for a proton of rigidity 445 MV (which means 100 MeV energy). These panels correspond to a maximum of solar activity reached in August 1999.}
    \label{fig:config_max}
\end{figure}

Figure \ref{fig:config_max} shows the 2D meridional and equatorial cuts at respectively $\phi=0$ and $\theta=\pi/2$ of the parallel (on the left) and perpendicular (on the right) mfp in AU for the maximum of activity. Here the configuration is very multipolar, with more than 2 current sheets visible in the meridional cut. The equatorial plane also shows 4 current sheets crossing the equatorial plane, but with less angular spread than at minimum of activity. Depending on the position of the Earth along its orbit at 1 AU, both the parallel and perpendicular diffusion are thus very different at maximum of activity, with favored axis of diffusion.

\begin{figure}[!t]
    \centering
    \subfigure{\includegraphics[width=0.46\columnwidth]{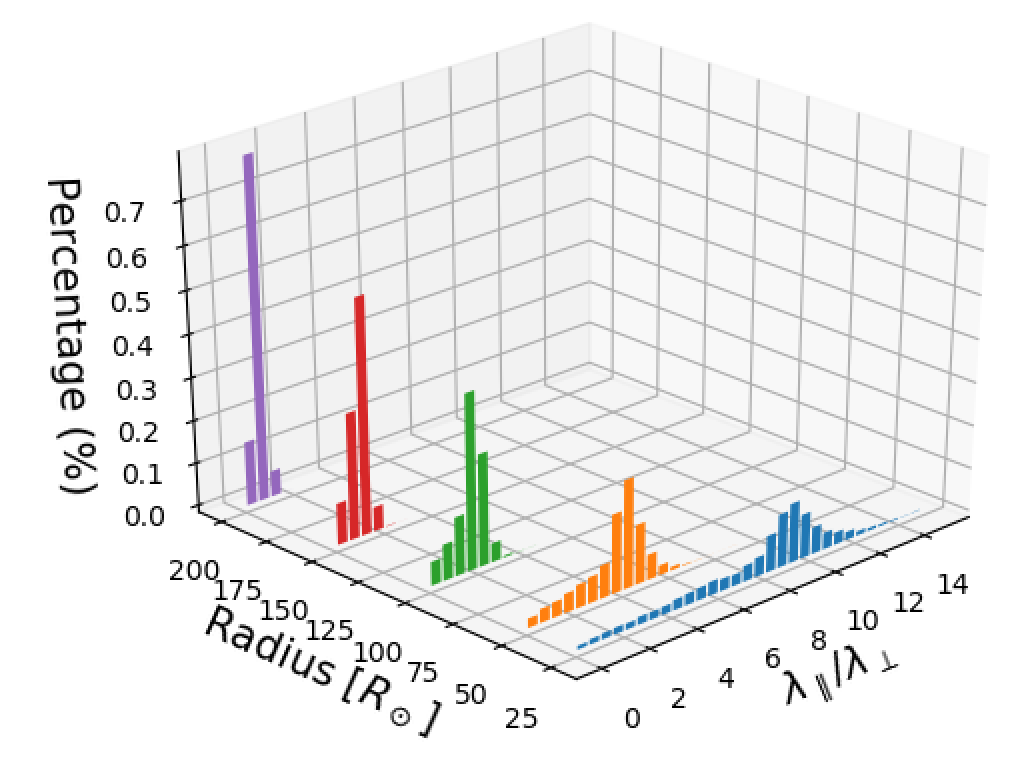}}
    \subfigure{\includegraphics[width=0.45\columnwidth]{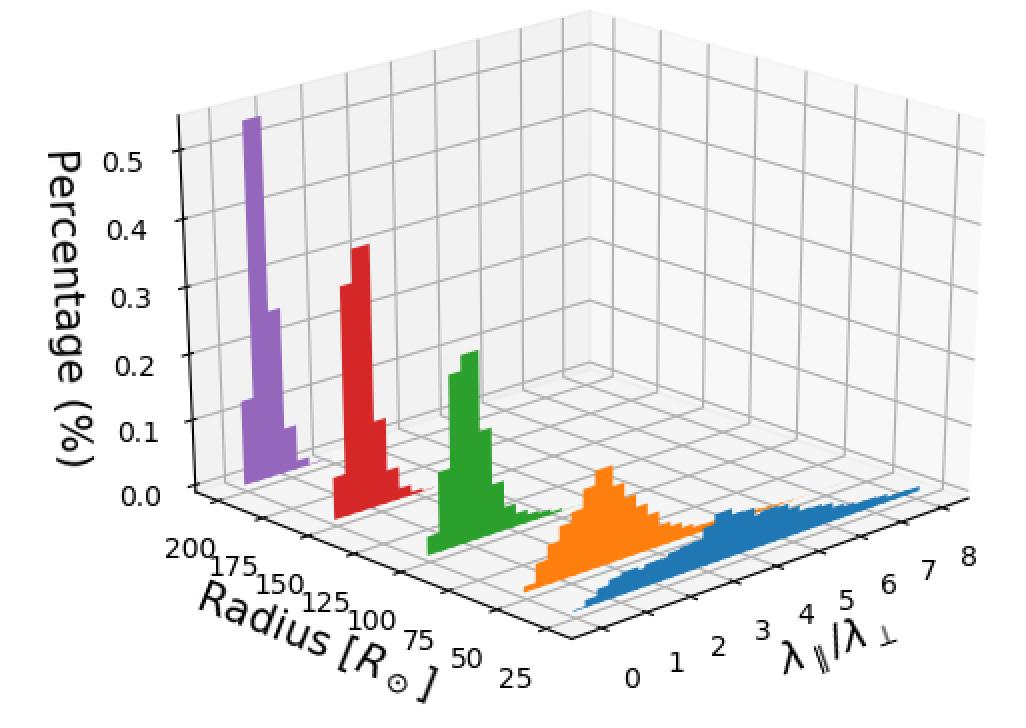}}
    \caption{Histograms of the ratio of the parallel and perpendicular mfp in AU for a proton of rigidity 445 MV (which means 100 MeV energy) in minimum of activity (on the left) and maximum of activity (on the right) at $r=50, 100, 150, 200 \ R_\odot$. The histograms show the distribution in percentage on a spherical shell at the corresponding radius.}
    \label{fig:hist_min_max}
\end{figure}

Figure \ref{fig:hist_min_max} shows 1D histograms of the distribution of values of the ratio of $\lambda_\parallel$ over $\lambda_\perp$ in spherical shells at $r=25, 50, 100, 150, 200 \ R_\odot$ to show the profile of diffusion at a given radius. The minimum of activity is on the left, the maximum on the right. As said before, a general trend observed is that the further we go from the Sun, the more this ratio diminishes, meaning that the perpendicular mfp becomes more and more important with the distance to the Sun. For the minimum of activity for example, at 25 $R_\odot$, the peak value is at 10, while at 1 AU, the peak value is around 2. At minimum of activity we see a drift for the peak value of the distribution : at $25R_\odot$, the peak is at 10, at $50R_\odot$ it is at 5, at $100R_\odot$ at 4, at $150R_\odot$ at 3 and at $200R_\odot$ at 2. The spread of the distribution is also different : the closer we are to the Sun, the more spread is the distribution with the peak reaching only 15\% of the total distribution; the further we go from the Sun, the more peaked is the distribution with more than 70\% of the values between 1 and 2 at 1 AU. This means that at minimum, there is a distinction of behavior close and far from the Sun : close to the Sun perpendicular mfp is smaller than parallel mfp but with a more irregular distribution of values. This means that the diffusion for SEPs and GCRs is different at minimum, because SEPs originate from the Sun and GCRs from outside the solar system. Such insights in the low corona can soon be correlated with the results from Parker Solar Probe to better understand the diffusion of SEPs in the corona. At maximum of activity, we observe the same behavior, except that the ratio $\lambda_\parallel/\lambda_\perp$ is smaller, with the peak of the distribution reaching only 4 at $25 R_\odot$. The distribution is also slightly more spread, with the peak of the distribution corresponding to only 50\% of the total distribution at 1 AU. This means that at maximum, because of the multiple current sheets, it is more probable to have a high perpendicular diffusion, and thus it is almost as important as the parallel one 25\% of the time at 1 AU.

\section{Conclusions}
\label{sec:concl}

In this study, we combine theory and numerical simulations to better understand the impact of magnetic field on the propagation of CRs between the Sun and the Earth. We use a 3D MHD polytropic wind simulation based on the PLUTO code, which can use either analytical formulations or data from synoptic maps to initialize the bottom boundary condition magnetic field at the surface of the star. This provides us realistic inputs for the heliospheric magnetic field and wind speed. We complete this simulation with approximations and fits for turbulence parameters such as $\lambda_s$ and $\epsilon$. We then use analytical formulations in post-processing to compute the parallel and perpendicular mfps associated with a proton of a certain rigidity or energy. For the parallel mfp, we used the formulation of \cite{zank_radial_1998}; for the perpendicular mfp, we used the formulation of \cite{shalchi_analytic_2004} derived from the work of \cite{bieber_nonlinear_2004}.

As the solar cycle evolves in time, the magnetic field is going from weak and mostly dipolar at minimum of activity, to strong and mostly quadrupolar at maximum of activity. We decoupled these two effects to study separately the influence of the amplitude and the geometry of the magnetic field on CR diffusion. We have thus run 3 cases : case D1 with a weak dipole, case D10 with a strong dipole and case Q1 with a weak quadrupole. The comparison between cases D1 and D10 shows that the magnetic field amplitude affects the amplitude of the diffusion coefficients : increasing $B_*$ by a factor 10 enhances $\lambda_\parallel$ of a factor 10 and decreases $\lambda_\perp$ by a factor 100. It also affects the spread of the current sheet, which means that the mfps present variations on a wider zone for a stronger field; hence the amplitude also affects the radial distribution of the mfps. However numerical effects can also enhance this tendency, so this result must be treated carefully. The comparison between cases D1 and Q1 shows that the magnetic field geometry does not affect the amplitude of the mfps but changes the location of the current sheet, which changes the latitudinal distribution of the mfps with more variations at the equator for the dipole, and more at $\theta=60\degree$ for the quadrupole. Finally we separated SEPs and GCRs by changing the energy of the particle. Our tests on cases D1 and D10 shows that the rigidity/energy dependency of the diffusion is actually very sensitive to the magnetic field amplitude. GCRs have higher values of mfps than SEPs with a more spread distribution of values. We have thus demonstrated that amplitude and geometry have a different impact on SEPs and GCRs, which could be a first step for retrieving information about the past of the solar-terrestrial interactions using CRs readings over the last centuries \citep{finley_solar_2019}.

We then applied our post-processing to configurations computed using a synoptic map. We simulated the state of the inner heliosphere within Earth orbit corresponding to the minimum of activity of October 1995 and the maximum of activity of August 1999. This allows us to have for the first time 3D maps of the CR diffusion between the Sun and the Earth at specific dates. This shows that the diffusion is highly non-axisymmetric for real configurations. We also show that for real configurations, there are different behaviors close and far from the Sun (below and above $150R_\odot$), especially at minimum of activity. We also show that perpendicular diffusion is not to be neglected at 1 AU, especially at maximum of activity when the current sheet configuration can be very complex. Thanks to this study, we thus provide useful and easy tools to derive CR diffusion from any wind simulation.

This study is a first step towards bridging theory and simulations of wind and CRs. We focused here on the diffusion coefficient, because it is one of the most difficult term to model in the Parker CR transport equation \citep{parker_passage_1965} ; it still needs to be completed with the drift coefficient to take into account the full influence of the magnetic field. We plan however to go beyond by modeling the other terms of the equation, which is rather straight-forward using the wind and magnetic field parameters provided by our 3D MHD simulation, and finally solving the Fokker-Planck equation. Thus we would be able to have the spatial distribution of CRs in the heliosphere to compare it with data from the Earth neutron monitors \citep{heber_cosmic_2006}, from the Moon data \citep{poluianov_solar_2018} or even Venus \citep{lorenz_gamma_2015} or Mars data \citep{lee_helium_2006}. In particular it would be interesting to add the effects of adiabatic cooling as it has a major influence on the modulation of CRs \citep{jokipii_effects_1979}. We can also change the population of particles with our formulations, seeing the effect of positive or negative charge, in particular in regard to the polarity of the magnetic field, in order to see the influence of the 22-year cycle on the modulations of CRs \citep{heber_cosmic_2006}. We also plan to apply such formulations to a more realistic wind model, because as we have explained it, the modeling of the turbulence presented here has limits; it is suited to have the turbulence evolving with the MHD model. Such work for a turbulence-based wind model with Alfvén wave heating is currently being undertaken, see \cite{reville_role_2020}. CRs propagation through the heliosphere, and especially in the lower corona, is bound to be an important subject in the years to come thanks to the combined efforts of Parker Solar Probe and Solar Orbiter; here we focused on the magnetic field, but the wind structures can also locally affect the propagation of CRs \citep{cohen_energetic_2020, mccomas_probing_2019}. Finally, to fully understand the complete dynamic of GCRs, we would need to extend our model to go beyond 1 AU, which could be done using more ressourceful numerical methods such as AMR (Adaptative Mesh Refinement). For space weather purposes, the model would need to be adapted to be time-dependent like in \cite{kim_predicting_2020}.

\begin{acknowledgements}
We thank Steve Tobias for the original motivation to perform this study. We thank Karl-Ludwig Klein and Sophie Masson for useful discussions. This work was supported by a CEA 'Thèse Phare' grant, by CNRS and INSU/PNST program, by CNES SHM funds and by the ERC Synergy grant WholeSun. Computations were carried out using CEA CCRT and CNRS IDRIS facilities within the GENCI 20410133 allocation, and a local meso-computer founded by DIM ACAV+. We thank Eric Buchlin and the MEDOC facility for hosting the simulation outputs.
\end{acknowledgements}


\bibliography{swsc}
\bibliographystyle{swsc}
   

\end{document}